\documentstyle[11pt]{article}
\normalbaselines

\newtheorem{thm}{Theorem}[section]
\newtheorem{lem}{Lemma}[section]
\newtheorem{prop}{Proposition}[section]

\begin{document}
\bibliographystyle{unsrt}

\def\bea*{\begin{eqnarray*}}
\def\eea*{\end{eqnarray*}}
\def\ba{\begin{array}}
\def\ea{\end{array}}
\count1=1
\def\be{\ifnum \count1=0 $$ \else \begin{equation}\fi}
\def\ee{\ifnum\count1=0 $$ \else \end{equation}\fi}
\def\ele(#1){\ifnum\count1=0 \eqno({\bf #1}) $$ \else \label{#1}\end{equation}\fi}
\def\req(#1){\ifnum\count1=0 {\bf #1}\else \ref{#1}\fi}
\def\bea(#1){\ifnum \count1=0   $$ \begin{array}{#1}
\else \begin{equation} \begin{array}{#1} \fi}
\def\eea{\ifnum \count1=0 \end{array} $$
\else  \end{array}\end{equation}\fi}
\def\elea(#1){\ifnum \count1=0 \end{array}\label{#1}\eqno({\bf #1}) $$
\else\end{array}\label{#1}\end{equation}\fi}
\def\cit(#1){
\ifnum\count1=0 {\bf #1} \cite{#1} \else 
\cite{#1}\fi}
\def\bibit(#1){\ifnum\count1=0 \bibitem{#1} [#1    ] \else \bibitem{#1}\fi}
\def\ds{\displaystyle}
\def\hb{\hfill\break}
\def\comment#1{\hb {***** {\em #1} *****}\hb }

\newcommand{\TZ}{\hbox{\bf T}}
\newcommand{\MZ}{\hbox{\bf M}}
\newcommand{\ZZ}{\hbox{\bf Z}}
\newcommand{\NZ}{\hbox{\bf N}}
\newcommand{\RZ}{\hbox{\bf R}}
\newcommand{\CZ}{\,\hbox{\bf C}}
\newcommand{\PZ}{\hbox{\bf P}}
\newcommand{\QZ}{\hbox{\rm eight}}
\newcommand{\HZ}{\hbox{\bf H}}
\newcommand{\EZ}{\hbox{\bf E}}
\newcommand{\GZ}{\,\hbox{\bf G}}

\font\germ=eufm10
\def\goth#1{\hbox{\germ #1}}
\vbox{\vspace{38mm}}

\begin{center}
{\LARGE \bf The Q-operator for Root-of-Unity Symmetry in Six Vertex Model}\\[10 mm] 
Shi-shyr Roan \\
{\it Institute of Mathematics \\
Academia Sinica \\  Taipei , Taiwan \\
(email: maroan@gate.sinica.edu.tw ) } \\[25mm]
\end{center}

\begin{abstract}
We construct the explicit $Q$-operator incorporated with the $sl_2$-loop-algebra symmetry of the six-vertex model at roots of unity. The functional relations involving the $Q$-operator, the six-vertex transfer matrix and fusion matrices are derived from the Bethe equation, parallel to the Onsager-algebra-symmetry discussion in the superintegrable $N$-state chiral Potts model. We show that the whole set of functional equations is valid for the $Q$-operator. Direct calculations in certain cases are also given here for clearer illustration about the nature of the $Q$-operator in the symmetry study of root-of-unity six-vertex model from the functional-relation aspect. 
\end{abstract}
\par \vspace{5mm} \noindent
{\it 1999 PACS}:  05.50.+q, 02.20.Tw, 75.10Jm \par \noindent

\setcounter{section}{0}
\section{Introduction}
\setcounter{equation}{0}
The aim of this paper is to construct the $Q$-matrix 
for the six-vertex model at roots of unity incorporable to the $sl_2$-loop-algebra symmetry found in \cite{DFM}, and the functional relations proposed in \cite{FM04, R05b}. Since initiated by Bethe in 1931 \cite{Be}, the free energy of the six-vertex model and the spectrum of the XXZ-spin chain with the periodic boundary condition,
\be
H_{\rm XXZ} = \frac{-1}{2} \sum_{\ell =1}^L ( \sigma_\ell^1 \sigma_{\ell+1}^1 +
\sigma_\ell^2 \sigma_{\ell+1}^2 + \triangle \sigma_\ell^3 \sigma_{\ell+1}^3 ), 
\ele(XXZ)
where $\sigma^i_\ell$ is the Pauli spin matrix at the site $\ell$ with $L+1=1$, have long been studied by means of Bethe ansatz. The results obtained from Bethe ansatz were shown by Baxter in the seventies to follow from the $TQ$-relation method by introducing an auxiliary '$Q$'-operator, a genuine invention in his solution of the eight-vertex model \cite{B71, B72, B73, Bax}. The $TQ$-relation method enables one to study the eigenvalues  of the transfer matrix $T$ without knowing the eigenvectors, but instead to work on the eigenvalues of the $Q$-operator. In this manner, there exist many $Q$-matrices which satisfy the $TQ$-equation (see, e.g. \cite{Bax, Kor} and references therein), but only a certain special one is expected to exhibit the additional symmetry appeared in the six-vertex model when the anisotropic parameter $q$ with $\triangle = \frac{1}{2} (q+q^{-1})$ is a root of unity. In this work, we find a such $Q$-operator. We demonstrate first by direct calculations in examples,  then by means of mathematical arguments in general cases, that the $Q$-operator we have constructed here is the one for the symmetry study of the root-of-unity six-vertex model, much in the same way as the chiral Potts transfer matrix in the Onsager-algebra symmetry study of the superintegrable $N$-state chiral Potts model (CPM) in \cite{R05o}.

In their study of the root-of-unity symmetry in eight-vertex model \cite{FM04}, Fabricius and McCoy proposed the conjectural functional relations for the eight-vertex transfer matrix $T$ and $Q_{72}$-matrix in \cite{B72}, analogous to the set of functional equations known in the $N$-state CPM  \cite{BBP}. After the occurrence of $\tau^{(2)}$-degeneracy in CPM was found to appear only in the superintegrable case \cite{R04}, the Fabricius-McCoy comparison between the root-of-unity eight-vertex model and superintegrable CPM was further analysed, and their common mathematical structure led to the discovery of the Onsager-algebra symmetry of $\tau^{(j)}$-model in the superintegrable CPM \cite{R05o}. These exact results in CPM can  serve as a valuable scheme in the study of symmetry problems of solvable lattice models, among which the root-of-unity six-vertex model is a distinguished one, due to the importance of the related XXZ-spin chain (\req(XXZ)), and the progress made on the newly found $sl_2$-loop-algebra symmetry in roots of unity case \cite{DFM, De05, FM00, FM001, FM01}. Indeed, the Fabricius-McCoy comparison in the limiting case of eight-vertex model with the vanishing elliptic nome  strongly suggested that a unified theory exists for both the root-of-unity symmetry of six-vertex model and the Onsager-algebra symmetry of CPM from the functional relation aspect, which was clearly spelled out in \cite{R05b}. Among the functional relations, the $QQ$-relation, equivalent to the $Q$-functional equation, encodes the essential features reflecting the symmetry of the model. In the theory of CPM, the $QQ$-relation (called the $T\widehat{T}$-relation in \cite{BBP}) played the vital role in the derivation of the whole set of functional equations, by which the solution of eigenvalue spectrum of the CPM transfer matrix, considered as the $Q$-operator, was obtained \cite{B90, MR}. Works in \cite{AMP, B93, B94} about the $Q$-eigenvalues in the superintegrable CPM has paved the way to both the qualitative and quantitative understanding of the degeneracy and Onsager-algebra symmetry of $\tau^{(j)}$-models \cite{R05o}. These results encourage us to view the CPM as a useful 'toy'-model to illustrate the symmetry nature of the root-of-unity six-vertex model, and hopefully to other models. For the purpose of displaying the common symmetry characters governing various models we will consider in this scheme, handling the precise formulation which correctly and uniformly presents the functional relations for all those models is a non-trivial problem. By adjusting some suitable normalizing factors to the original operators appeared in the CPM functional relations in \cite{BBP}, we refine the quantity to extract a (somewhat unpleasant or more complicated) representation of functional relations, which can be extended successfully to include both the root-of-unity six- and eight-vertex models so that the symmetry property of the $Q$-operator will be clearly presented. Furthermore,
we also hope the effort of understanding the common integrable structure of those models will help to provide some clues to solutions of certain unsolved important problems in CPM, such as the CPM correlation functions, parallel to the results known for the 
correlation functions of the six- and eight-vertex models. In the context of the $\tau^{(2)}$-model where the six-vertex model is with a particular field (see \cite{B049}, page 3), Bazhanov and Stroganov showed that the column transfer matrix of the $L$-operator   ((2.19) of \cite{BazS})) possesses the properties of Baxter's $Q$-matrix with features of symmetry operators similar to, but not exactly the same as, the $sl_2$-loop algebra which plays the role of symmetry in the 'zero-field' 
six-vertex model (\cite{B04} Sect. 3). Indeed, the Onsager-algebra symmetry operators inherited from the Baxter's $Q$-matrix provide the precise description about the symmetry structure of $\tau^{(j)}$-model in the superintegrable CPM \cite{R05o}. In the study of 'zero-field' six-vertex model, which we loosely call the six-vertex model in this paper, the explicit form of the six-vertex $Q$-operator which displays the root-of-unity symmetry is our main concern. This $Q$-operator will behave like the $Q_{72}$-operator in the root-of-unity eight-vertex discussion in \cite{FM04}, which was invented by Baxter in 1972 for his study of eigenvalues of the eight-vertex transfer matrix \cite{B72}.
In this work, we produce a $Q$-operator of the six-vertex model at roots of unity by following Baxter's method of constructing $Q_{72}$ in \cite{B72}, and demonstrate that this special $Q$-operator is indeed in accordance with the root-of-unity symmetry of six-vertex model from the functional-relation aspect. Note that in the process of constructing this six-vertex $Q$-operator, we first produce certain operators $Q_R, Q_L$, then $Q$ by a normalized factor as in the eight-vertex case \cite{B72}. The $Q_R, Q_L$ here cannot be reached from the eight-vertex $Q_R$- and $Q_L$-operator by taking the vanishing elliptic nome limit. However in cases when $Q_{72}$ is explicitly known, the limit exists. It should be interesting to study the comparison of the $Q$-operator in this paper with a six-vertex limit of the eight-vertex $Q_{72}$ because the conclusion about the $Q$-operator in other cases not discussed in this work depends on it. Indeed the $Q$-operator we obtain in this paper coincides with Baxter's classical result on the six-vertex limit (for a generic $q$ and $S^z=0$) of the eight-vertex $Q$-operator in \cite{B73} (formula (101)).

This paper is organized as follows. In Sect. \ref{sec: 6v}, we summarize the main features of the transfer matrix $T(z)$ and $sl_2$-loop-algebra symmetry of the six-vertex model at roots of unity for the use of later discussions. 
In Sect. \ref{sec.FR}, we review the functional relations in the study of Onsager-algebra symmetry of superintegrable CPM and the loop-algebra symmetry of the root-of-unity six-vertex model in \cite{R05o, R05b}. Through the identical mathematical structure about the Bethe equation of these two models  related to the evaluation polynomial in describing the $T^{(2)}$-degeneracy, we derive the set of functional equations, based on the known scheme in the superintegrable CPM case \cite{R05o}, and then extended to the root-of-unity six-vertex model. The $TQ$-, fusion, and $T^{(j)}Q$-relation are the established relations, discussed in subsection \ref{ssec. TQfu}; while the $QQ$- and $Q$-functional equations are the symmetry constraints of the models, examined in subsection \ref{ssec.QQ}. Indeed, we give a more elaborate discussion there about the $QQ$- and $Q$-functional equation of the root-of-unity six-vertex model for later uses. Sect. \ref{sec.6Q} contains the main result of this paper about the $Q$-operator of the root-of-unity six-vertex model. First in subsection \ref{ssec. Q}, by imitating Baxter's method in \cite{B72} of producing  $Q_{72}$-operator in the eight-vertex model at roots of unity, we construct a $Q$-operator, depending the variable $z^{\frac{1}{2}}$ for the $TQ$-relation, and commuting with the spin inversion operator and  spatial translation. Then 
we make direct, but non-trivial, checks for the cases of size $L \leq 6$ in subsection \ref{ssec. L<6} to verify the $Q$-functional relation by using the explicit form of the $Q$-operator. The computational evidence has revealed the characteristics of the $Q$-operator, and further enhanced its correct role
in the symmetry study of the six-vertex model at roots of unity. In subsection \ref{ssec. funL}, we show  by rigorous mathematical arguments that the functional relations for a general $L$ are valid for the $Q$-operator obtained in this paper.   We close in Sect. \ref{sec. F} with some concluding remarks. 

{\bf Notations.} In this
paper, $\ZZ, \CZ$ will denote 
the ring of integers, complex numbers respectively, $\ZZ_N = \ZZ/N\ZZ$  and ${\rm i} = \sqrt{-1}$.  
For a positive integers $n, L$, we denote
by $\stackrel{L}{\otimes} \CZ^n$ the tensor
product of $L$-copies of the vector space $\CZ^n$.

\section{Six-vertex Model at Roots of Unity and Loop-algebra Symmetry \label{sec: 6v}}
\setcounter{equation}{0}
We start with some basic notions in the six-vertex model at roots of unity considered in the later sections. This summary will be rather sketchy, but also serve to establish the notation: more detailed information can be found in any of the standard references listed in the bibliography, such as \cite{DFM, De05, FM01} and references therein. 

It is known that the $L$-operator\footnote{Here we use the convention in \cite{FM01}, Eq. (1.3): $a={\rm i} \sinh \frac{1}{2}(v- {\rm i} \gamma)$, $b=- {\rm i} \sinh \frac{1}{2}(v + {\rm i} \gamma)$, $c= - {\rm i} \sinh {\rm i} \gamma$, with the variables $z= -e^v , q= -e^{{\rm i} \gamma}$. Note that the $q$ here differs from that in \cite{FM01} by minus sign for its connection with $U_q(\widehat{sl_2})$ as  Eq. (2.3) in \cite{De05}, where $-e^{-z}$ is equal to $z^{\frac{1}{2}}$ here.}
\be
L (z^{\frac{1}{2}})  =  \left( \begin{array}{cc}
        L_{0,0}  & L_{0,1}\\
        L_{1,0}& L_{1,1} 
\end{array} \right) = \left( \begin{array}{cc}
    z^{\frac{1}{2}} q^{\frac{-\sigma^z }{2}} - z^{\frac{-1}{2}} q^{\frac{\sigma^z }{2}}, & (q-q^{-1}) \sigma_- \\
(q-q^{-1}) \sigma_+ , & z^{\frac{1}{2}} q^{\frac{\sigma^z }{2}} - z^{\frac{-1}{2}} q^{\frac{- \sigma^z }{2}}
\end{array} \right)  , \ \ z \in \CZ , 
\ele(6VL)
is the Yang-Baxter solution for the $R$-matrix 
$$
R (z) = \left( \begin{array}{cccc}
        z^{\frac{-1}{2}} q - z^{\frac{1}{2}} q^{-1}  & 0 & 0 & 0 \\
        0 &z^{\frac{-1}{2}} - z^{\frac{1}{2}} & q-q^{-1} &  0 \\ 
        0 & q-q^{-1} &z^{\frac{-1}{2}} - z^{\frac{1}{2}} & 0 \\
     0 & 0 &0 & z^{\frac{-1}{2}} q - z^{\frac{1}{2}} q^{-1} 
\end{array} \right).
$$
Using the $L$-operator, one defines the monodromy matrix, $\bigotimes_{\ell=1}^L L_\ell (z^{\frac{1}{2}})$ with $L_\ell (z^{\frac{1}{2}})= L (z^{\frac{1}{2}})$ at the site $\ell$, and its trace  
\be
T (z) = {\rm tr}_{aux} ( \bigotimes_{\ell=1}^L L_\ell (z^{\frac{1}{2}})), \ \ {\rm for} \ z \in \CZ ,
\ele(6Tz)
form a commuting family of $(\stackrel{L}{\otimes} \CZ^2)$-operators, which define the transfer matrix  (of size $L$ with the periodic condition) of the six-vertex model.
The XXZ Hamiltonian (\req(XXZ)) is the logarithmic $z\frac{d}{dz}$-derivative of $T (z)$ at $z= q^{-1}$, where $\triangle = \frac{1}{2}(q+q^{-1})$.
The transfer matrix $T(z)$ commutes with the $z$-component of total spin $S^z$, the spin-inversion operator $R$,
$$
S^z = \frac{1}{2} \sum_{\ell=1}^L \sigma_\ell^z , \ \ \ \ \ \ R = \bigotimes_{\ell} \sigma_\ell^{x},
$$
and  the spatial translation operator $S_R$, which takes the $j$th column to $(j+1)$th one for $1 \leq j \leq L$ with the identification $L+1 = 1$. In the basis $\otimes | \pm 1 \rangle_\ell$ for the diagonal $\sigma_\ell^z$ where $\sigma^z | \pm 1 \rangle = \pm | \pm 1\rangle$, the number of down spins $n$ is known to relate to $S^z$ by $n= \frac{L}{2} -S^z$, and $(-1)^n$ is the quantum number for the operator $S = \bigotimes_{\ell} \sigma_\ell^{z}$, which commutes with  $T(z)$. 

When $q$ is a $N$th root of unity, degenerate multiplets for the spectrum of XXZ-eigenvalues occur with the $sl_2$-loop-algebra symmetry for the eigenspaces \cite{DFM}. 
In later sections, we are going to examine this loop-algebra symmetry of the transfer matrix $T(z)$ through the $Q$-operator and related functional relations, as in the symmetry study of superintegrable CPM.  
For definiteness, in this paper, we will confine the discussion of the six-vertex model only to the cases for {\it even} $L$ (the chain-size),  and  {\it odd} $N$  (the order of $q$). We also assume all $q$, $\omega (:= q^2)$, $q^{\frac{1}{2}}$ are primitive $N$th roots of unity.  {\it Except section $\ref{sec.FR}$}, the discussion of the transfer matrix $T(z)$ and all relevant operators will be confined only to sectors $S^z \equiv 0 \pmod{N}$, unless stated otherwise, i.e., they are regarded as operators of $V$ where
\be
V : = \sum_{ \sum_\ell \beta_\ell  \equiv 0 \pmod{N} } \CZ \otimes_\ell | \beta_\ell \rangle , \ \ \beta_\ell = \pm 1 .
\ele(V)
Then the $sl_2$-loop-algebra symmetry is generated by $\frac{2S^z}{N}$ and the operators 
\bea(ll)
S^{\pm (N)} =& \sum_{1 \leq j_1 < \cdots < j_N \leq L} 1 \otimes \cdots \otimes 1 \otimes \sigma_{j_1}^\pm \otimes q^{-\sigma^z}\otimes \cdots \otimes  q^{-\sigma^z} \\
&\otimes \sigma_{j_2}^\pm \otimes q^{-2\sigma^z} \cdots  \otimes q^{-(N-1) \sigma^z} \otimes \sigma_{j_N}^\pm \otimes 1 \cdots \otimes 1 , \\
T^{\pm (N)} =& \sum_{1 \leq j_1 < \cdots < j_N \leq L} 1 \otimes \cdots \otimes 1 \otimes \sigma_{j_1}^\pm \otimes q^{\sigma^z}\otimes \cdots \otimes  q^{\sigma^z} \\
&\otimes \sigma_{j_2}^\pm \otimes q^{2\sigma^z} \cdots  \otimes q^{(N-1) \sigma^z} \otimes \sigma_{j_N}^\pm \otimes 1 \cdots \otimes 1 , 
\elea(ST)
with the relation $T^{\pm (N)} = R S^{\mp (N)} R^{-1}$ \cite{DFM, De05}.

\section{Functional Relations of Superintegrable CPM and Root-of-unity Six-vertex Model \label{sec.FR} }
\setcounter{equation}{0}
Here we give a summary of functional relations in \cite{R05b}, which govern both cases of the superintegrable CPM and the six-vertex model at roots of unity, and a more detailed account of $QQ$- and $Q$-functional relation in the six-vertex model case will be discussed in subsection \ref{ssec.QQ} for later uses. In the discussion of this section, the spin $S^z$ of the six-vertex model can be arbitrary, i.e., no the $N$-multiple restriction for $S^z$; while in the CPM case, the order $N$ is not required to be odd, nor the chain-length $L$ being even, and $\omega$ is a $N$th root of unity. 

We begin our consideration by reviewing the Onsager-algebra symmetry of superintegrable CPM in \cite{R05o, R05b}. 
The rapidities  of $N$-state CMP are represented by coordinates  of the genus-$(N^3-2N^2+1)$ curve 
\begin{eqnarray*}
{\goth W}  : & \left\{ \begin{array}{l}
ka^N + k'c^N = d^N , \\
kb^N + k'd^N = c^N  \end{array}  \right.   \Longleftrightarrow  
\left\{ \begin{array}{l}
k x^N  = 1 -  k'\mu^{-N}  , \\
k y^N  = 1 -  k' \mu^N  \end{array}  \right.
\end{eqnarray*}
where $[a, b, c, d] \in \PZ^3$, $(x, y, \mu)=(\frac{a}{d}, \frac{b}{c}, \frac{d}{c}) \in \CZ^3$,  and $k'$ is the parameter with $k^2= 1 - k'^2 \neq 0, 1$. Then the variables $t:= xy, \lambda:= \mu^N$ define the hyperelliptic curve of genus $N-1$:
\be
 t^N = \frac{(1- k' \lambda )( 1 - k' \lambda^{-1}) }{k^2 }. 
\ele(CPhpE) 
Consider the following symmetries of rapidities of order $N$ and 2:
\bea(lll)
U :  [a,b,c,d] \mapsto [\omega a,b,c,d], & \Longleftrightarrow   (x, y, \mu) \mapsto ( \omega x, y, \mu), & \Longleftrightarrow (t, \lambda) \mapsto (\omega t, \lambda) ;  \\
C :  [a,b,c,d] \mapsto [b,a, d, c ], & \Longleftrightarrow (x, y, \mu) \mapsto ( y, x , \mu^{-1} ), & \Longleftrightarrow (t, \lambda) \mapsto (t, \lambda^{-1}) .
\elea(UC)
The $\tau^{(2)}$-model is constructed from Yang-Baxter solutions $G_p(t)$ for $p=[a, b, c, d] \in {\goth W}$: 
$$
 b^2 G_p (t)  = \left( \begin{array}{cc}
       b^2 - t d^2 X & ( bc - \omega a d X )  Z   \\
       -t (b c - a d X ) Z^{-1} & - t c^2  + \omega  a^2  X  
\end{array} \right)  \ , \ \ \ t \in \CZ \ ,
$$
associated to an inhomogeneous $R$-matrix of the six-vertex model \cite{BazS, R04}, where $X, Z$ are the Weyl operator of $\CZ^N$:  $X |n \rangle = |n +1 \rangle$, $ Z |n \rangle = \omega^n |n \rangle $ for $n \in \ZZ_N$. The $\tau^{(2)}_p$-matrix is defined by  
$$
\tau^{(2)}_p (t) := {\rm tr}_{aux} ( \bigotimes_{\ell=1}^L G_{p, \ell} ( \omega t)), \ \ \ \ G_{p, \ell} (t) =  G_p (t) \  {\rm  at \  site} \ \ell .
$$
The '$Q$'-operator associated to $\tau^{(2)}$-matrix is 
the CPM transfer matrix \cite{BBP, BazS}, which is the $(\stackrel{L}{\otimes} \CZ^N)$-operator defined by 
\begin{eqnarray*}
T_{\rm cp} (p; s)_{\sigma_1 , \ldots , \sigma_L}^{\sigma_1' , \ldots , \sigma_L'} = \prod_{l=1}^L
\overline{W}_{p,s}(\sigma_l - \sigma_l')
W_{p,s}(\sigma_l - \sigma_{l+1}') \ , \ \ \sigma_l, \sigma_l' \in
\ZZ_N ,
\end{eqnarray*}
where $p, s \in {\goth W}$, and $W_{p,s}$, $W_{p,s}$ are Boltzmann weights defined by the $N$-cyclic vectors:  $\frac{W_{p,s}(n)}{W_{p,s}(0)}  = \prod_{j=1}^n
\frac{d_pb_s-a_pc_s \omega^j}{b_pd_s-c_pa_s\omega^j}$, $
\frac{\overline{W}_{p,s}(n)}{\overline{W}_{p,s}(0)} 
 = \prod_{j=1}^n
\frac{\omega a_pd_s-
d_pa_s\omega^j}{ c_pb_s- b_pc_s \omega^j}$, which satisfy the well-known 
star-triangle relation. Then for a fixed $p \in {\goth W} $, $\{ T_{\rm cp} (p; s) \}_{s \in {\goth W}} $ form a commuting family of operators, commuting with the  spin-shift operator $X(:= \prod_\ell X_\ell) $ of $(\stackrel{L}{\otimes} \CZ^N)$ and the spatial translation $S_R$. In the theory of CPM, there is a collection of functional relations among the CPM-transfer-matrix $T_{\rm cp}(p; *)$, the $\tau^{(2)}_p$- and fusion $\tau^{(j)}_p$-matrices,  involving the rapidity-symmetries $U, C$ in (\req(UC)). In this paper, the discussion of CPM will be confined only to the superintegrable case, i.e., the rapidity $p$ given by
$\mu_p = 1$, $ x_p = y_p = \eta^{\frac{1}{2}}$ with $\eta := ( \frac{1-k'}{1+k'})^{\frac{1}{N}}$, where simplification occurs for the functional relations. For simplicity, the label $p$ will be omitted for operators at the superintegrable point $p$, and simply write  $\tau^{(2)}(*)$, $T_{\rm cp}(*)$,  etc. It is convenient to use the normalized CPM transfer matrix\footnote{The $Q$-operator here differs from the ${\sc Q}_{cp}$ in \cite{R05o} by a scale factor: $(\eta^{\frac{-1}{2}}x_s)^{P_a}(\eta^{\frac{-1}{2}}y_s)^{P_b}\mu_s^{-P_\mu} Q(s) = {\sc Q}_{cp}(s)$.}:
\be
Q (s) = \frac{T_{\rm cp} (s)(1- \eta^{\frac{-N}{2}} x_s^N)^L}{ N^L (1 - \eta^{\frac{-1}{2}} x_s )^L(\eta^{\frac{-1}{2}}x_s)^{P_a}(\eta^{\frac{-1}{2}}y_s)^{P_b}\mu_s^{-P_\mu}} \ ( =: e^{-{\rm i}P} \widehat{Q} (s) ),
\ele(CPQ)
where $P_a, P_b, P_\mu$ are quantum numbers in the CPM theory and $P$ is the total moment.  Introduce  the variable $\tilde{t} = \eta^{-1} t_s$ and write $\tau^{(2)}$-matrix by $
\widetilde{\tau}^{(2)}(\tilde{t}) = \tau^{(2)}( t_s)$. By explicit results about eigenvalues of CPM transfer matrix in \cite{AMP, B93, B94}, the Onsager-algebra symmetry of $\tau^{(j)}$-model was verified rigorously in \cite{R05o} through the functional relations. Furthermore, if we normalize the $\tau^{(2)}$- and $\tau^{(j)}$-matrices by 
\be
T^{(2)}(\tilde{t}):= \frac{\omega^{-P_b} (1- \widetilde{t}^N)^L \widetilde{\tau}^{(2)}(\omega^{-1} \widetilde{t}) }{(1- \omega^{-1} \widetilde{t})^L(1-  \widetilde{t})^L}, \ \ T^{(j)}( \widetilde{t}) = \frac{\omega^{-(j-1)P_b} (1- \widetilde{t}^N)^L \widetilde{\tau}^{(j)}(\omega^{-1}\widetilde{t}) }{\prod_{k=-1}^{j-2}(1- \omega^k \widetilde{t})^L} ,
\ele(CPTj)  
the $TQ$-, and fusion relations of CPM possess the identical structure as those in the root-of-unity six-vertex model through the evaluation polynomial of the respective symmetry algebra. By these common structures, as a theory parallel to CPM, we propose in \cite{R05b} the functional relations for the $Q$-operator of the six-vertex model to display the root-of-unity symmetry, now described below in a uniform manner.

\subsection{Bethe equation, $TQ$-relation, and fusion relation \label{ssec. TQfu}}
With the spectral variable $t_s, z$ in CPM and six-vertex model, respectively, we denote  
$$
H(\widetilde{t}) = \left \{ \begin{array}{lll} 
  \frac{1 - \widetilde{t}^N}{1-\widetilde{t}},  &\widetilde{t}:= \eta^{-1} t_s, & {\rm in \ CPM} , \\
  1-\widetilde{t}, & \widetilde{t}:= q z, &{\rm in \ six}-{\rm vertex \ model } . \end{array} \right.
$$
For non-zero parameters $v_1, \cdots, v_m$, distinct and not containing a complete $N$-cyclic string (which means $\{ \omega^j v \}_{j \in \ZZ_N}$ for some $v \neq 0$),  we consider the non-zero polynomial 
$$
F(\widetilde{t}) =  \prod_{i=1}^m (1 + \omega^{\frac{-n_0}{2}} v_i \widetilde{t}), 
$$
where $n_0$ is the integer defined by $n_0:= -2, 1 $ for  CPM and the six-vertex model, respectively. Note that as before, $\omega = q^2$ in the six-vertex model case.  
For an integer $r$ with $ 0 \leq r \leq N-1$, and the polynomial $F(\widetilde{t})$, we define the function 
\be
P(\widetilde{t})=  \sum_{j=0}^{N-1} \frac{H ( \omega^j \widetilde{t})^L (\omega^j \widetilde{t})^{ -r}}{F(\omega^j \widetilde{t}) F(\omega^{j+1} \widetilde{t})}.
\ele(P)
$P(\widetilde{t})$ is invariant when changing $\widetilde{t}$ to $\omega \widetilde{t}$, hence depends only on $\widetilde{t}^N$. The polynomial criterion for $P(\widetilde{t})$ is that the $v_i$'s in $F(\widetilde{t})$ satisfy the relations
\be
\frac{ H(- \omega^{\frac{n_0}{2}-1} v_i^{-1} )^L}{H(- \omega^{\frac{n_0}{2}} v_i^{-1})^L} 
= - \omega^{-r} 
\frac{F(- \omega^{\frac{n_0}{2}-1} v_i^{-1}) }{F(- \omega^{\frac{n_0}{2}+1} v_i^{-1})} ,
\ \ \ i = 1, \ldots, m .
\ele(Bethe)
Indeed the above equation is the Bethe equation for the superintegrable CPM ((4.4) in \cite{AMP}, (6.22) in \cite{B93}) and for the six-vertex model where $r \equiv \frac{L}{2}-m \ (=|S^z|) \pmod{N}$ with $0 \leq r \leq N-1$. Then the polynomial $P(\widetilde{t})$ in (\req(P)) is the evaluation polynomial for the symmetry algebra of the model \cite{ FM01, R05o}\footnote{The evaluation function of six-vertex model in \cite{FM01} is given by (3.9) with $Y(v) = \sum_{j = 0}^{N-1} \overline{a}(v +2 (j+1) {\rm i}\gamma  )$, where $\overline{a}(v)=   \frac{\sinh^L \frac{1}{2}(v-{\rm i}\gamma)}{\prod_{i =1}^m \sinh \frac{1}{2}(v_i -v) \sinh \frac{1}{2}(v_i - v + 2 {\rm i} \gamma)}$ in (2.47). In terms of variables $z, \widetilde{t}$ and Bethe roots $v_i$s here, $\overline{a}(v)
= 2^{2m-L} (\prod_{i=1}^m v_i)(q^{-1}z)^{\frac{-L}{2}+m}  \frac{(1- q^{-1}z)^L}{\prod_{i=1}^m (1+v_i z )(1+ \omega^{-1} v_i z)}$, which implies $Y(v) = 2^{2m-L} (\prod_{i=1}^m v_i )  \widetilde{t}^{\frac{-L}{2}+m +r} P(\widetilde{t})$.}. 

Based on the common feature of the Bethe equation (\req(Bethe)), and known functional relations in CPM  about operators $T^{(j)}(\widetilde{t})$, $Q(s)$ in 
(\req(CPTj)) (\req(CPQ)) and symmetries $U, C$ in (\req(UC)), we now derive functional relations for both CPM and the root-of-unity six-vertex model. The $T^{(2)}(\widetilde{t})$ and symmetries $U, C$ in six-vertex model are defined by
\bea(rl)
T^{(2)}(\widetilde{t}) = & z^{\frac{L}{2}} T(z) , \ \ ( \widetilde{t}:= q z) , \\
U: z \mapsto \omega z , &
C: s \mapsto -s , \ \ ( s:= z^{\frac{1}{2}}) ,
\elea(6TUC)
where $T(z)$ is the transfer matrix (\req(6Tz)). Note that $T^{(2)}(\widetilde{t})$ is a $\widetilde{t}$-polynomial operator, and in terms of the variable $s$, $U$ is defined by $U(s)= q s$. 

By the Bethe equation (\req(Bethe)), the $T^{(2)}$-eigenpolynomial, denoted again by $T^{(2)}(\widetilde{t})$ (when no confusion could arise), satisfies the relation
\be
T^{(2)}( \widetilde{t})  F( \widetilde{t}) = \omega^{-r} H(\widetilde{t})^L F( \omega^{-1}\widetilde{t})  +  H ( \omega^{-1} \widetilde{t})^L F( \omega \widetilde{t}).    
\ele(TF)
Define the $T^{(j)}$-operators for non-negative $j$ recursively by setting $T^{(0)}( \widetilde{t}) = 0$, $T^{(1)}( \widetilde{t}) = H( \omega^{-1}\widetilde{t})^L$, and the fusion relation
\be 
T^{(j)}( \widetilde{t}) T^{(2)}(\omega^{j-1} \widetilde{t}) = \omega^{-r} H(\omega^{j-1} \widetilde{t})^L T^{(j-1)}( \widetilde{t})  + H(\omega^{j-2}\widetilde{t})^L T^{(j+1)}( \widetilde{t}),  \ \ j \geq 1 . 
\ele(Tfus)  
Express $T^{(j+1)}$ in   $T^{(j)}$ and $T^{(j-1)}$ from the above relations; then by induction argument, one obtains the eigenpolynomial expression of $T^{(j)}(\widetilde{t})$: 
\be
T^{(j)}(\widetilde{t}) = F( \omega^{-1} \widetilde{t}) F( \omega^{j-1}\widetilde{t}) \sum_{k=0}^{j-1} \frac{ H(\omega^{k-1}\widetilde{t})^L \omega^{-kr} }{F(\omega^{k-1}\widetilde{t})F(\omega^k \widetilde{t})} \ , \ \ \ \ j \geq 1 .
\ele(TjF)
By this, follows the '$N$-periodic' relation of $T^{(j)}$'s:
$$
T^{(N+j)}(\widetilde{t}) =   \omega^{-r}\frac{F( \omega^{j-1}\widetilde{t})}{F(\widetilde{t})} T^{(N-1)}(\omega \widetilde{t}) + T^{(j)}(\widetilde{t})+\frac{ H(\omega^{-1}\widetilde{t})^L F( \omega^{j-1}\widetilde{t}) }{F( \widetilde{t})}, \ \ j \geq 0 ,
$$
in particular, the case $j=1$ yields the boundary fusion relation 
\be
T^{(N+1)}(\widetilde{t}) = \omega^{-r} T^{(N-1)}( \omega \widetilde{t})
  + 2 H ( \omega^{-1} \widetilde{t})^L . 
\ele(TfBdy)
Note that the fusion relations, (\req(Tfus)) and (\req(TfBdy)), are equivalent to those for $\tau^{(j)}$'s in the superintegrable CPM case ((53) in \cite{R05o}). 

We now discuss the $Q$-operator, i.e. a family of  non-degenerated commuting operators $Q(s)$ depending on the variable $s$ (which is algebraically related to $\widetilde{t}$  by (\req(CPhpE)) or (\req(6TUC))),  such that $[T(\widetilde{t}), Q(s)] =0$ with the following $TQ$-relation:
\be
T^{(2)}( \widetilde{t})  Q (s) = \omega^{-r} H(\widetilde{t})^L Q (U^{-1}s) +  H ( \omega^{-1} \widetilde{t})^L Q (Us) ,  
\ele(T2Q)
where $U$ is the $s$-symmetry in (\req(UC)) or (\req(6TUC)). Note that by comparing (\req(TF)) with (\req(T2Q)), the $\widetilde{t}$-polynomial $F(\widetilde{t})$ is expected to be a factor of an eigenvalue of $Q (s)$. 
Using (\req(T2Q)) and (\req(Tfus)), one finds the following $T^{(j)}Q$-relation by induction argument for $j \geq 0 $ with $T^{(0)}=0$,
\be
T^{(j)}( \widetilde{t})
= Q(U^{-1} s) 
Q(U^{j-1} s)  \sum_{k =0}^{j-1} \bigg( \omega^{-k r} H( \omega^{k -1} \widetilde{t})^L  
Q(U^{k -1} s)^{-1}  Q(U^k s)^{-1}  \bigg). 
\ele(TjQ)
Indeed for a $Q$-operator in (\req(T2Q)), the $T^{(j)}Q$-relation (\req(TjQ)) is equivalent to the fusion relations, (\req(Tfus)) and (\req(TfBdy)). 
The $TQ$- and $T^{(j)}Q$-relations here are equivalent to the $\widetilde{\tau}^{(2)}{\sc Q}_{cp}$- and $\widetilde{\tau}^{(j)}{\sc Q}_{cp}$-relations in CPM case ((54) (55) in \cite{R05o}). In the case of six-vertex model, the $TQ$-relation (\req(T2Q)) is valid for any $Q$-operator (up to a suitable normalizing factor), so is the $T^{(j)}Q$-relation (\req(TjQ)).

\subsection{The $QQ$-relation and $Q$-functional relation \label{ssec.QQ}}
Now we discuss the constraint on the $Q$-operator to encode the essential features about the symmetry of CPM and the root-of-unity six-vertex model. First we note that by (\req(TjF)), the $T^{(j)}$-eigenpolynomials and evaluation polynomial $P(\widetilde{t})$ in (\req(P)) are related by 
\bea(cl)
T^{(j)}(\widetilde{t}) + \omega^{-jr} T^{(N-j)}(\omega^j\widetilde{t}) = (\omega^{-1} \widetilde{t})^r F( \omega^{-1} \widetilde{t}) F( \omega^{j-1}\widetilde{t}) P(\omega^{-1} \widetilde{t}), & 0 \leq j \leq N , \\
T^{(N)}(\widetilde{t}) = (\omega^{-1} \widetilde{t})^r F( \omega^{-1} \widetilde{t})^2  P(\omega^{-1} \widetilde{t}),& 
\elea(TTp)
which in CPM case are equivalent to the relations between $\tau^{(j)}$-eigenvalues and the evaluation polynomial ((73) in \cite{R05o}). We expect the $Q$-operator in accord with (\req(TTp)) and the expression of $P(\widetilde{t})$ in (\req(P)), which leads to the following constraint of $Q (s)$:
\begin{eqnarray*}
&T^{(j)}(\widetilde{t}) + \omega^{-jr} T^{(N-j)}(\omega^j\widetilde{t}) = \widetilde{t}^r Q( U^{-1}s) Q( U^{j-1}s) \sum_{k=0}^{N-1} \frac{H(\omega^{k-1}\widetilde{t})^L (\omega^k\widetilde{t})^{-r}}{ Q(U^{k-1}s)Q(U^k s)}, \ \ 0 \leq j \leq N , \\
&T^{(N)}(\widetilde{t}) = \widetilde{t}^r Q( U^{-1} s)^2  \sum_{k=0}^{N-1} \frac{H(\omega^{k-1}\widetilde{t})^L (\omega^k\widetilde{t})^{-r}}{ Q(U^{k-1}s)Q(U^k s)}.
\end{eqnarray*}
The $U$-invariant operator $\sum_{k=0}^{N-1} \frac{H(\omega^{k-1}\widetilde{t})^L (\omega^k\widetilde{t})^{-r}}{ Q(U^{k-1}s)Q(U^k s)}$ has the  eigenvalue  $\omega^{-r} P(\omega^{-1} \widetilde{t})$. We now use the symmetry $C$ in (\req(UC)) or (\req(6TUC)) to state the  functional relation required for $Q (s)$:
\be
P(\widetilde{t}_0) F(\widetilde{t}_0)^2  Q(Cs) =  Q(s) \sum_{k=0}^{N-1} H(\omega^{k-1}\widetilde{t})^L (\omega^k\widetilde{t})^{-r} Q(U^{k-1}s)^{-1}Q(U^k s)^{-1} ,
\ele(Qg) 
with $\widetilde{t}_0 = 1, 0$ in CPM and the six-vertex model, respectively. Here, we abuse the notation by using $P(\widetilde{t}_0) F(\widetilde{t}_0)^2$ to denote the diagonal operator acting as the $(P(\widetilde{t}_0) F(\widetilde{t}_0)^2)$-multiple on the $T^{(2)}$-eigenspace with the $T^{(2)}$-eigenvalue determined by $F(\widetilde{t})$.   Then, the above constraint relations of $T^{(j)}$ and $Q$ yield the following $QQ$-relation:
\bea(cl)
P(\widetilde{t}_0) F(\widetilde{t}_0)^2 \widetilde{t}^r Q(U^{-1} s) Q(U^{j-1}C s )   =  T^{(j)}(\widetilde{t})   + \omega^{-jr}T^{(N-j)} (\omega^j \widetilde{t})  , & 0 \leq j \leq N ,  \\
P(\widetilde{t}_0) F(\widetilde{t}_0)^2 \widetilde{t}^r Q(U^{-1} s) Q(U^{-1}Cs )  =  T^{(N)}(\widetilde{t}).
\elea(QQ)
In the case of CPM, (\req(Qg)) and (\req(QQ)) are valid, and equivalent to the ${\sc Q}_{cp}$-functional and  ${\sc Q}_{cp}\widehat{\sc Q}_{cp}$-relation in \cite{R05b} (for formulas (33) (56) by using  the equality $N^L \omega^{P_b} e^{{\rm i}P}= P(1) F(1)^2$). 
Indeed, relations in (\req(Qg)) and (\req(QQ)) are all equivalent by the following result.
\begin{thm}\label{thm:QQ-Q} 
For an integer $j$ between $0$ and $N$, the $j$th $QQ$-relation in $(\req(QQ))$ is equivalent to the Q-functional relation $(\req(Qg))$, hence equivalent to the whole set of relations $(\req(QQ))$.
\end{thm}
{\it Proof.} 
By (\req(TjQ)), one has
$$
\begin{array}{ll}
\omega^{-jr} T^{(N-j)}(\omega^j\widetilde{t}) &=  Q( U^{j-1} s) Q( U^{-1} s) \sum_{k=0}^{N-j-1} \frac{ H(\omega^{k+j-1}\widetilde{t})^L \omega^{-(k+j)r} }{Q(U^{k+j-1}s)Q(U^{k +j} s)} .
\end{array}
$$
By substituting $T^{(j)}$ of (\req(TjQ)) and the above expression into $(\req(QQ))$, follows  
the equivalence of the $j$th relation in $(\req(QQ))$ and the $Q$-functional relation (\req(Qg)). $\Box$ \par \vspace{.2in} 

For the rest of this paper, we consider only the root-of-unity six-vertex model in sectors $S^z \equiv 0 \pmod{N}$, where $0 = r \equiv \frac{L}{2}-m$. The polynomial $F(\widetilde{t})$ and Bethe equation (\req(Bethe)) are now expressed by
\be
F(\widetilde{t}) =  \prod_{i=1}^m (1 +  v_i z), \ \ \ \ \ 
\frac{ (v_i + q^{-1}  )^L}{(v_i + q )^L} 
 = - \prod_{l=1}^m \frac{ v_i -  \omega^{-1} v_l }{v_i -  \omega v_l  } 
\ ~ \ ~ {\rm for} \ \ i = 1, \ldots, m ,
\ele(6FBe0)
with $\widetilde{t} = q z$.
Hence, $F(0)=1$, and $P(0)=N$. Note that the equation (\req(6FBe0)) is unaltered when replacing $v_i$ by $v_i^{-1}$. Denote $F^\dagger(\widetilde{t}) =  \prod_{i=1}^m (1 +  v_i^{-1} z)$. Then the polynomials $P(\widetilde{t})$,$P^{\dagger}(\widetilde{t})$ of (\req(P)) for  $F(\widetilde{t})$, $F^\dagger(\widetilde{t})$ resp. satisfy the following 'reciprocal' relation:
$$
\widetilde{t}^{L-2m} P( \widetilde{t}^{-1}) = (\prod_{i=1}^m v_i)^{-2} P^\dagger (\widetilde{t}).
$$   
By this and $P^\dagger (0) \neq 0$, the $\widetilde{t}$-degree of the polynomial $P(\widetilde{t})$ is given by
\be
{\rm deg} P (\widetilde{t})= L-2m .
\ele(PtN)
The $T^{(2)}$-eigenvalue is determined by the Bethe polynomial $F(\widetilde{t})$ via (\req(TF)), now in the form
\be
T^{(2)}( \widetilde{t})  F( \widetilde{t}) =  (1-\widetilde{t})^L F( \omega^{-1}\widetilde{t})  +  (1- \omega^{-1} \widetilde{t})^L F( \omega \widetilde{t}).    
\ele(TF0)
The non-degeneracy of $T^{(2)}( \widetilde{t})$, i.e.  a $T^{(2)}$-eigenvalue with  one-dimensional eigenspace, is equivalent to the constant function for the polynomial $P (\widetilde{t})$ associated to $F (\widetilde{t})$. As a consequence of (\req(PtN)), we have the following characterization of non-degenerated $T^{(2)}$-eigenvalues:
\begin{lem}\label{lem:ndT} 
Let $T^{(2)}( \widetilde{t})$ be a $T^{(2)}$-eigenvalue corresponding to a Bethe polynomial $F (\widetilde{t})$ in $(\req(6FBe0))$. Then $T^{(2)}( \widetilde{t})$ is non-degenerated if and only if the degree $F (\widetilde{t})$ is equal to $\frac{L}{2}$.
\end{lem} $\Box$ \par \vspace{.1in} \noindent
Now the $T^{(2)}Q$-, $QQ$- and $Q$-functional relations, (\req(T2Q)) (\req(QQ)) and  (\req(Qg)), take the form:
\begin{eqnarray}
&T^{(2)}( \widetilde{t})  Q (s) =  (1-\widetilde{t})^L Q (q^{-1}s) +  (1- \omega^{-1} \widetilde{t})^L Q (qs) , \ s:= z^{\frac{1}{2}} , \label{6TQ0}\\
& N Q(q^{-1} s) Q(-q^{j-1} s )   =  T^{(j)}(\widetilde{t})   +  T^{(N-j)} (\omega^j \widetilde{t})  , &  0 \leq j \leq N , \label{QQ0} \\
 &N Q(-s) =  Q(s) \sum_{k=0}^{N-1} (1- \omega^{k-1}\widetilde{t})^L  Q(q^{k-1}s)^{-1}Q(q^k s)^{-1} .
\label{6Q0} 
\end{eqnarray}
In the next section, we produce a $Q$-operator with the above properties.

\section{The Q-operator for Six-vertex Model at Roots of Unity \label{sec.6Q}}
\setcounter{equation}{0}
We first construct the $Q$-operator of the root-of-unity six-vertex model in subsection \ref{ssec. Q} by following Baxter's method of producing the $Q$-operator of the eight-vertex model in Appendix C of \cite{B72}. Then in subsection \ref{ssec. L<6} we verify the $Q$-functional equation (\ref{6Q0})  for the size $L \leq 6$ by direct computations. Finally, in subsection \ref{ssec. funL} we show the functional relations hold for a general $L$ by using the explicit forms of the $Q$-operator and fusion matrices.

\subsection{Construction of the $Q$-operator \label {ssec. Q} }
The $L$-operator (\req(6VL)) is the matrix with the $\CZ^2$-auxiliary space and the following $\CZ^2$-(quantum-space) operator-entries: 
$$
\begin{array}{l}
L_{0,0} = \left( \begin{array}{cc}
        a  & 0\\
        0& b 
\end{array} \right), \ L_{0,1} = \left( \begin{array}{cc}
        0  & 0\\
        c& 0
\end{array} \right) , \ L_{1,0} = \left( \begin{array}{cc}
        0  & c\\
        0& 0 
\end{array} \right) , \ L_{1,1} = \left( \begin{array}{cc}
        b  & 0\\
        0& a 
\end{array} \right), 
\end{array}
$$
where 
\be
a=a(z^{\frac{1}{2}})= z^{\frac{1}{2}} q^{\frac{-1 }{2}} - z^{\frac{-1}{2}} q^{\frac{1 }{2}}, \ \ b= b (z^{\frac{1}{2}})= z^{\frac{1}{2}} q^{\frac{1 }{2}} - z^{\frac{-1}{2}} q^{\frac{-1 }{2}}, \ \ c= q-q^{-1}. 
\ele(6abc)
Since the chain-size $L$ is even, the entries of the transfer matrix $T(z)$ are $z$-functions. Hence $T(z)$ is unaltered by changing the variable $z^{\frac{1}{2}}$ to $-z^{\frac{1}{2}}$ in the $L$-operator, i.e., $T(z) = {\rm tr}_{aux}  \bigotimes_{\ell=1}^L L_\ell (-z^{\frac{1}{2}})$.\footnote{The functions,  $a(-z^{\frac{1}{2}})$, $b(-z^{\frac{1}{2}})$ and $c$, correspond to Boltzmann weights  of the six vertex model in the face-model description in \cite{B04} (22):
$W_{6v}(\alpha,\beta,\gamma, \delta) = \delta(\alpha,\gamma) (zq)^{\frac{\beta-\alpha-\gamma+\delta}{4}}a(-z^{\frac{1}{2}}) - \delta (\beta, \delta) (zq^{-1})^{\frac{\beta-\alpha-\gamma+\delta}{4}} b(-z^{\frac{1}{2}}) .
$} 

Consider an ${\sf S }$-operator with $\CZ^N$-auxiliary space and $\CZ^2$-quantum space, ${\sf S }= ({\sf S }_{i,j})_{i, j \in \ZZ_N} $, such that the $\CZ^2$-operators ${\sf S }_{i,j}$ are of the form
\be
{\sf S}_{i,j} = {\sf v}_{i, j } \tau_{i, j}, \ \ \ \ \ {\sf v}_{i, j } = \left(\begin{array}{c}
        v_1  \\
        v_2 
\end{array} \right), \ \  \tau_{i, j}= (\tau_1, \tau_2), 
\ele(Sij)
where $v_j$'s depend on the $z$-variable, and $\tau_j$'s are constants in $\CZ$. 
The general form of the matrix ${\sf Q}_R$ we shall use is 
$$
{\sf Q}_R= {\rm tr}_{\CZ^N} ( \bigotimes_{\ell =1}^L {\sf S}_{\ell}), \ \ \ {\sf S}_{\ell}= {\sf S} \ {\rm at \ the \ site} \ \ell,
$$
with $T{\sf Q}_R = {\rm tr}_{\CZ^2 \otimes \CZ^N} ( \bigotimes_{\ell =1}^L {\sf U}_{\ell})$ and  ${\sf U}_{\ell}= {\sf U}$ at the site $\ell$, where ${\sf U}$ is the following matrix  associated to the $L$- and ${\sf S}$-operator with $\CZ^2 \otimes \CZ^N $-auxiliary space and $\CZ^2$-quantum space:
$$
{\sf U} = \left( \begin{array}{cc}
        L_{0,0} {\sf S } & L_{0,1} {\sf S} \\
        L_{1,0} {\sf S} & L_{1,1} {\sf S }
\end{array} \right) .
$$
The operator $T{\sf Q}_R$ will decompose into the sum of two matrices if we can find a $2N$ by $2N$ scalar matrix ${\sf M}$ (independent of $z$)
such that 
$$
{\sf M}^{-1} {\sf U} {\sf M} = \left( \begin{array}{cc}
        {\sf A}  & {\sf B}  \\ 
         0 & {\sf D}
\end{array} \right) .
$$
As in \cite{B72}, the above required form is unaffected by postmultiplying ${\sf M}$ by a upper blocktriangular matrix; together with a similar transformation of ${\sf S}$, we can in general choose
$$ 
{\sf M} = \left( \begin{array}{cc}
        I_N  & 0 \\
        \delta & I_N 
\end{array} \right) , \ \ \delta = {\rm dia} [\delta_0, \cdots, \delta_{N-1}] .
$$
Hence,
$$
{\sf M}^{-1} {\sf U} {\sf M} = \left( \begin{array}{cc}
        L_{0,0} {\sf S } +  L_{0,1} {\sf S }\delta,  & L_{0,1} {\sf S}  \\ 
       - \delta L_{0,0} {\sf S } + L_{1,0} {\sf S } - \delta L_{0,1} {\sf S } \delta  +
       L_{1,1} {\sf S} \delta , & - \delta L_{0,1} {\sf S }  + L_{1,1} {\sf S }
\end{array} \right) ,
$$
and the condition on ${\sf S}_{i,j}$'s in (\req(Sij)) for the vanishing lower blocktriangular matrix  is 
$$
\left( \begin{array}{cc}
        -a   \delta_i  + b  \delta_j &  c    \\
        -c \delta_i \delta_j & a  \delta_j -b   \delta_i 
\end{array} \right) \left( \begin{array}{c}
        v_1  \\
        v_2 
\end{array} \right) (\tau_1, \tau_2) = 0 .
$$
For non-zero ${\sf S}_{i,j}$, this in turn yields
$$
0= (a^2+b^2-c^2) \delta_i \delta_j - ab (\delta_i^2 + \delta_j^2)= (z+z^{-1} - (q + q^{-1})) ( \delta_i - q \delta_j )(\delta_i - q^{-1}\delta_j),  
$$
hence $\delta_i = q^{\pm 1} \delta_j $ and 
$$
v_1:v_2 = \left \{\begin{array}{ll} c : ( a q  -b )  \delta_j  & {\rm if} \ \delta_i = q \delta_j , \\ 
c : (aq^{-1} - b) \delta_j &{\rm if} \ \delta_i = q^{-1}\delta_j .
\end{array} \right.
$$
We choose the $L$-operator to be $L(- z^{\frac{1}{2}})$ in the above argument without affecting $T(z)$, i.e., replacing $a, b$ by $a(- z^{\frac{1}{2}}), b(- z^{\frac{1}{2}})$, and then set 
\be
\delta = {\rm dia}[1, q, \cdots, q^{N-1}], \ \ \ {\rm i.e.} \ \delta_j = q^j \ \ {\rm for} \ j =0, \ldots N-1 . 
\ele(diaQ)
Then ${\sf S}_{i, j}$ is equal to zero except $j-i = \pm 1 \in \ZZ_N$, in which case the two-vector ratio for ${\sf v}_{i, j }$ in (\req(Sij)) is given by $v_1: v_2 = 1: z^{\frac{j-i}{2}} q^{\frac{i-j}{2}} \delta_j$.  Hence, we can set 
\be
{\sf v}_{i, j} = \left\{\begin{array}{ll}
 ((zq)^{\frac{i-j}{4}} q^{\frac{-i}{2}},  (zq)^{\frac{j-i}{4}} q^{\frac{i}{2}} )^t &{\rm if} \ j- i= \pm 1 , \\
0 & {\rm otherwise} , \end{array}  \right. 
\ele(vij)
with the ${\sf S}$-operator in the form 
$$
{\sf S} = \left( \begin{array}{cccccc}
        0  & {\sf S}_{0,1} & 0&\cdots   & 0&{\sf S}_{0,N-1}\\
        {\sf S}_{1,0} &  0& {\sf S}_{1,2} & \ddots & &\vdots \\
0&&\ddots&\ddots&& \\
\vdots&&\ddots&\ddots&&0 \\
0&&&\ddots& &{\sf S}_{N-2,N-1}\\
{\sf S}_{N-1,0} &0&\cdots &0&{\sf S}_{N-1,N-2}&0
\end{array} \right) .
$$
One can show the $(i,j)$-th entry of the diagonal block matrices of ${\sf M}^{-1} {\sf U} {\sf M}$ satisfies the relations
$$
\begin{array}{l}
(L_{0,0} {\sf S })_{i,j}(z) +  (L_{0,1} {\sf S })_{i,j}(z) \delta_j=  a (- z^{\frac{1}{2}})q^{\frac{j-i}{2}} {\sf S}_{i,j} ( z \omega ), \\
(- \delta L_{0,1} {\sf S })_{i, j}(z)  + (L_{1,1} {\sf S })_{i, j}(z) = b(- z^{\frac{1}{2}}) q^{\frac{-(j-i)}{2}} {\sf S}_{i,j} ( z \omega^{-1} ) ,
\end{array}
$$
which, by $L$ even, imply  
\be
T(z) {\sf Q}_R (z) = b(z^{\frac{1}{2}})^L  {\sf Q}_R ( z \omega^{-1}) + a(z^{\frac{1}{2}})^L {\sf Q}_R ( z \omega) .
\ele(TQR)
We now replace ${\sf Q}_R$ by ${\sf Q}_L$, and form the product ${\sf Q}_LT$ by using ${\sf Q}_L= {\rm tr}_{\CZ^N} ( \bigotimes_{\ell =1}^L \widehat{\sf S}_{\ell})$ where the operator $\widehat{\sf S}= (\widehat{\sf S}_{i,j})_{i,j \in \ZZ_N}$ with $\widehat{\sf S}_{i,j}$ in the form
\be
\widehat{\sf S}_{i, j} = \widehat{\tau}_{i, j}\widehat{\sf v}_{i, j}, \ \ \ \ \ \widehat{\tau}_{i, j} = \left(\begin{array}{c}
        \hat{\tau}_1  \\
        \hat{\tau}_2 
\end{array} \right), \ \  \widehat{\sf v}_{i, j}= ( \hat{v}_1, \hat{v}_2).  
\ele(hSij)
We repeat the above working, replacing ${\sf S}$, $L(-z^{\frac{1}{2}})$ by $\widehat{\sf S}$, $L(z^{\frac{1}{2}})$, but using the same ${\sf M}$ with $\delta$ in (\req(diaQ)). We find that 
\be
\widehat{\sf v}_{i, j}  = \left\{ \begin{array}{ll}
 ( (zq)^{\frac{i-j}{4}} q^{\frac{j}{2}}, (zq)^{\frac{j-i}{4}} q^{\frac{-j}{2}} ) &{\rm if} \ j- i= \pm 1 , \\
0 & {\rm otherwise} ,
\end{array} \right.
\ele(hvij)
and 
\be
{\sf Q}_L (z) T(z)  = b(z^{\frac{1}{2}})^L  {\sf Q}_L ( z \omega^{-1}) + a(z^{\frac{1}{2}})^L {\sf Q}_L ( z \omega) .
\ele(TQL)
Note that the $z$-dependence of ${\sf Q}_R (z), {\sf Q}_L (z)$ in (\req(TQR)), (\req(TQL)) indeed means operator-functions of $z^{\frac{1}{2}}$. In order to construct a $Q$-operator commuting with $T$, as in \cite{B72} (C28) it suffices to show 
\be
{\sf Q}_L (w) {\sf Q}_R (z) = {\sf Q}_L (z) {\sf Q}_R (w). 
\ele(QLR)
To do this we follows the method in \cite{B04} (formulas (48) (49)) by considering the product functions of (\req(vij)) and (\req(hvij)), $f(w, z | i, j ; k, l) = \widehat{\sf v}_{i, j}(w) {\sf v}_{k, l} (z)$:
$$
f(w, z | i, j ; k, l) = \left\{ \begin{array}{ll}
(wq)^{\frac{j-i}{4}} (zq)^{\frac{l-k}{4}} q^{\frac{-j+k}{2}} + (wq)^{\frac{i-j}{4}} (zq)^{\frac{k-l}{4}} q^{\frac{j-k}{2}} &{\rm if} \ |j-i| =|l-k|=1 , \\
0 & {\rm otherwise} , \end{array}  \right.
$$
and looking for an auxiliary function $P( w, z | n)$ for $ n \in \ZZ_N$ such that  
\be 
f(z, w | i, j ; k, l) = P( w, z | k-i) f(w, z | i, j ; k, l) P( w, z | l-j)^{-1}.
\ele(fP)
Since the product ${\sf Q}_L (z) {\sf Q}_R (w)$ differs only the boundary contribution when interchanging $w$ by $v'$,   
$$
{\sf Q}_L (z) {\sf Q}_R (w) = P(w, z | k_1-i_1) {\sf Q}_L (w) {\sf Q}_R (z) P(w, z | k_{L+1}-i_{L+1})^{-1} ,
$$
this in turn implies the commutation relation (\req(QLR)) as the $P$-factors cancel out by the periodicity of boundary condition. There are four cases, $|j-i|=|l-k|=1$, to consider for the above function $P$. The relation (\req(fP)) automatically holds for $j-i= l-k$, and the other two cases yield just one condition for $P$: 
$$
\frac{P(w, z | n+1)}{P(w, z | n-1)} = \frac{(wq)^{\frac{1}{2}} + q^n (zq)^{\frac{1}{2}}}{ (zq)^{\frac{1}{2}} + q^n (wq)^{\frac{1}{2}}}.
$$
By $N$ odd and $q$ being an $N$th root of unity, the function $P$ is determined by the relation
$$
\frac{P(w,z | 2n)}{P(w, z | 0)} = \prod_{k=1}^n \frac{(wq)^{\frac{1}{2}} + q^k (zq)^{\frac{1}{2}}}{ (zq)^{\frac{1}{2}} + q^k (wq)^{\frac{1}{2}}} \ \ \ \ \ \ {\rm for } \ n \in \ZZ_N .
$$
By this, (\req(QLR)) holds. If we define 
\be
{\sf Q}(z) := {\sf Q}_R(z) {\sf Q}_R(z_0)^{-1} = {\sf Q}_L (z_0)^{-1} {\sf Q}_L(z) ,
\ele(Qdef)
then 
\be
[ T (z), T (w) ] = [ T (z), {\sf Q} (w) ] = [ {\sf Q} (z), {\sf Q} (w) ]= 0 , 
\ele(TQcom)
and 
\be
T(z){\sf Q} (z)   = b(z^{\frac{1}{2}})^L  {\sf Q} ( z \omega^{-1}) + a(z^{\frac{1}{2}})^L {\sf Q} ( z \omega) .
\ele(TQRL)

Note that ${\sf v}_{i, j}, \widehat{\sf v}_{i, j}$ in (\req(vij)), (\req(hvij)) satisfy the relation, $
q^{\frac{-\sigma^z}{2}}{\sf v}_{i, j} = {\sf v}_{i+1, j+1}$, $ \widehat{\sf v}_{i, j} = \widehat{\sf v}_{i+1, j+1} q^{\frac{-\sigma^z}{2}}$. It is convenient to choose $\tau_{i, j}, \widehat{\tau}_{i, j}$ in (\req(Sij)), (\req(hSij)) with the relations, $q^{\frac{\sigma^z}{2}}
\tau_{i, j}= \tau_{i+1, j+1} $, $q^{\frac{-\sigma^z}{2}} \widehat{\tau}_{i, j} =   \widehat{\tau}_{i+1, j+1}$. Then,  
$$
q^{\frac{-\sigma^z}{2}} {\sf S}_{i, j} q^{\frac{\sigma^z}{2}} = {\sf S}_{i+1, j+1}  ,
\ \
q^{\frac{\sigma^z}{2}} \widehat{\sf S}_{i, j}q^{\frac{-\sigma^z}{2}}  =   \widehat{\sf S}_{i+1, j+1} .
$$
So $\tau_{i,j}$'s are determined by $\tau_{0, 1}, \tau_{0,N-1}$; the same for $\widehat{\tau}_{i, j}$'s by $\widehat{\tau}_{0, 1}, \widehat{\tau}_{0, N-1}$. Since (\req(Qdef)) is unaffected by post- and pre-multiplying ${\sf Q}_R(z)$, ${\sf Q}_L(z)$ respectively by constant matrices, we may set 
$$
\tau_{0, 1}= \langle 1|, \ \ \tau_{0,N-1}= \langle -1| , \ \  \widehat{\tau}_{0, 1}= |1 \rangle, \ \ \widehat{\tau}_{0,N-1}= |-1  \rangle,  
$$
where $\langle \pm 1 |$ are the dual bases of $| \pm 1 \rangle$. Hence $\tau_{i, j}$'s and $\widehat{\tau}_{i, j}$'s are zeros except $j-i \equiv \pm 1 \pmod{N}$, in which cases one has the expression
$$
\tau_{i, j} = q^{\frac{i(j-i)}{2} } \langle j-i |, \ \ \ \widehat{\tau}_{i, j} = q^{\frac{-i(j-i)}{2} } | j-i \rangle, \ \ \ {\rm for} \ j-i \equiv \pm 1 .
$$
Then by (\req(vij)) and (\req(hvij)), the only non-zero ${\sf S}_{i, j}$ , $\widehat{\sf S}_{i, j}$ are  
\bea(ll)
{\sf S}_{i,j}(z)  =  q^{\frac{i(j-i)}{2} } \sum_{  \alpha =\pm 1}  q^{\frac{-i \alpha }{2}} (zq)^{\frac{-\alpha(j-i)}{4}}  | \alpha \rangle \langle j-i | , & {\rm for} \ j-i \equiv \pm 1 \pmod{N}, \\ 
\widehat{\sf S}_{i, j}(z) = q^{\frac{-i(j-i)}{2}} \sum_{  \beta =\pm 1}  q^{\frac{j \beta}{2}} (zq)^{\frac{-(j-i) \beta }{4}}  | j-i \rangle \langle \beta | ,&  {\rm for} \ j-i \equiv \pm 1 \pmod{N}.
\elea(Sfom)
Note that up to $(-1)$-power factors, the entries of ${\sf S}_{i,j}(z), \widehat{\sf S}_{i, j}(z)$ are the same as the weights in formulas (27) (43) of \cite{B04}: indeed by  $z= e^v, q=e^{\lambda}$, and the identification $\alpha, j, i$ with $b-a, c, d$, ${\sf S}_{i,j}(z)$ corresponds to $W_Q(v|a,b, c,d)$ in \cite{B04}; the same for $\widehat{\sf S}_{i, j}(z)$ and $\widehat{W}_Q(v|a,b, c,d)$ by identifying $i, j, \beta$ with $a, b, c-d$.

We now calculate the matrix form of ${\sf Q}_R$ and ${\sf Q}_L$. The relations, (\req(Sfom)) and ${\sf Q}_R  = \sum_{i_\ell} \otimes_{\ell=1}^L {\sf S}_{i_\ell, i_{\ell+1} }(z)$ with $i_\ell \in \ZZ_N$ and $i_{L+1} = i_1$, in turn yield 
$$
{\sf Q}_R = \sum_{i_\ell } \ ' \sum_{  \alpha_\ell =\pm 1} 
q^{\frac{\sum_\ell i_\ell (i_{\ell +1}-i_\ell )}{2} } q^{\frac{-\sum_\ell i_\ell \alpha_\ell }{2}} (zq)^{\frac{-\sum_\ell \alpha_\ell (i_{\ell +1} -i_\ell)}{4}}  \otimes_\ell | \alpha_\ell \rangle \otimes_\ell \langle i_{\ell+1} -i_\ell| ,
$$
where the prime-summation means the sum of those indices $i_\ell$'s with $i_{\ell+1} -i_\ell = \pm 1$. Denote $\beta_\ell = i_{\ell+1} -i_\ell = \pm 1$ for $1 \leq \ell \leq L$, and $i= i_1 \in \ZZ_N$. Then $\sum_{\ell}\beta_\ell \equiv i_{L+1} - i_1 = 0   \pmod {N}$, and ${\sf Q}_R $, as an operator of $V$ in (\req(V)), is expressed by 
$$
\begin{array}{ll}
{\sf Q}_R  &=  \sum_{  \alpha_\ell, \beta_\ell  =\pm 1}  
q^{\frac{\sum_\ell ( \beta_\ell- \alpha_\ell ) \sum_{s=1}^{\ell -1} \beta_s }{2} } (zq)^{\frac{-\sum_\ell \alpha_\ell \beta_\ell}{4}}  \sum_{i =0}^{N-1}  q^{\frac{i \sum_\ell (\beta_\ell -  \alpha_\ell )}{2} }    \otimes_\ell | \alpha_\ell \rangle \otimes_\ell \langle \beta_\ell  |, \\
&= N  \sum_{  \alpha_\ell, \beta_\ell  =\pm 1}  
q^{\frac{\sum_\ell ( \beta_\ell - \alpha_\ell ) \sum_{s=1}^{\ell -1} \beta_s }{2} } (zq)^{\frac{-\sum_\ell \alpha_\ell \beta_\ell}{4}}   \otimes_\ell | \alpha_\ell \rangle \otimes_\ell \langle \beta_\ell|.
\end{array}
$$
Since $\sum_{1 \leq s < \ell \leq L} \beta_s \beta_\ell
= \sum_{s} \beta_s \sum_{\ell  > s} \beta_\ell \equiv - \sum_s  \beta_s (\sum_{ \ell \leq s} \beta_\ell ) \equiv   -L - \sum_{1 \leq \ell < s \leq L} \beta_\ell \beta_s \pmod{N}$, one has
\be
\sum_\ell \beta_\ell  \sum_{s=1}^{\ell-1} \beta_s \equiv  \frac{-L}{2} \pmod{N},
\ele(sum) 
by which we obtain the matrix expression of ${\sf Q}_R$:
\be 
{\sf Q}_R = N q^{\frac{-L }{4} }  \sum_{  \alpha_\ell, \beta_\ell } 
q^{\frac{ - \sum_\ell  \alpha_\ell \sum_{s=1}^{\ell -1} \beta_s }{2} } (zq)^{\frac{-\sum_\ell \alpha_\ell \beta_\ell}{4}}   \otimes_\ell | \alpha_\ell \rangle \otimes_\ell \langle \beta_\ell| ,
\ele(QRm)
where indices $\alpha_\ell, \beta_\ell$ for $1 \leq \ell \leq L$ in the summation are $\pm 1$ with $\sum_{\ell} \alpha_\ell \equiv \sum_{\ell} \beta_\ell \equiv 0 \pmod{N}$. By $L$ even, the above expression of ${\sf Q}_R$ indeed defines an operator with $s$-polynomial entries, where as before $s= z^{\frac{1}{2}}$; and we shall also write ${\sf Q}_R= {\sf Q}_R (s)$. 
We repeat the above calculation, replacing ${\sf Q}_R, {\sf S}_{i, j}$ by ${\sf Q}_L, \widehat{\sf S}_{i, j}$, then we find the matrix form of the $V$-operator ${\sf Q}_L$:
\be 
{\sf Q}_L(s) = N q^{\frac{L}{4}} \sum_{\alpha_\ell, \beta_\ell } q^{\frac{ - \sum_\ell \alpha_\ell  \sum_{s=1}^{\ell-1} \beta_s}{2}} (zq)^{\frac{-\sum_\ell \alpha_\ell \beta_\ell }{4}}   \otimes_\ell | \alpha_\ell \rangle  \otimes_\ell  \langle \beta_\ell | .
\ele(QLm)
Hence ${\sf Q}_L$ and ${\sf Q}_R$ differ by a scalar, both commuting with the spin-inversion operator $R$,
\be
{\sf Q}_L(s) = q^{\frac{L}{2}} {\sf Q}_R(s), \ \ [R , {\sf Q}_R(s)] = [ R, {\sf Q}_L(s)] = 0 .
\ele(QLRR)
Using $\sum_{\ell} \alpha_\ell \equiv \sum_{\ell} \beta_\ell \equiv 0 \pmod{N}$, we find that the spatial translation commutes with ${\sf Q}_R, {\sf Q}_L$:
\be
[S_R , {\sf Q}_R(s)] = [ S_R, {\sf Q}_L(s)] = 0 .
\ele(SQ)

\subsection{Justification of the $Q$-functional relation for $L=2, 4, 6$ by direct computations \label{ssec. L<6}}
Here, we demonstrate by examples that 
the ${\sf Q}$-operator in (\req(Qdef)) satisfies the $QQ$- and $Q$-functional relation in subsection \ref{ssec.QQ}.  We modify ${\sf Q}_R, {\sf Q}_L$ by a normalized factor in accordance to the six-vertex transfer matrix $T^{(2)}$ in (\req(6TUC)), and write 
$$
Q_R (s) = z^{\frac{L}{4}} {\sf Q}_R (s), \ \ Q_L (s) = z^{\frac{L}{4}} {\sf Q}_L (s). 
$$
By (\req(QLRR)) and (\req(SQ)), the $s$-polynomial operators $Q_R, Q_L$  commuting with $R, S_R$  with $Q_L(s) = q^{\frac{L}{2}} Q_R(s)$. Denote 
$$
Q (s) :=  Q_R(s) Q_R(0)^{-1} = Q_L (0)^{-1} Q_L(s) . 
$$
As in (\req(TQcom)) and (\req(TQRL)), $Q(s)$ is the $Q$-operator associated to $T^{(2)}$ so that the $T^{(2)}Q$-relation (\ref{6TQ0}) holds.
By (\req(QRm)), $Q_R (0)= N q^{\frac{-L}{4}}{\rm Id}$, and the matrix expression of $Q(s)$ is given by
\be 
Q (s)  =  \sum_{  \alpha_\ell, \beta_\ell = \pm 1 } 
q^{\frac{ - \sum_\ell  \alpha_\ell \sum_{s=1}^{\ell -1} \beta_s }{2} } q^{\frac{-\sum_\ell \alpha_\ell \beta_\ell}{4}} s^{\frac{\sum_\ell (1-\alpha_\ell \beta_\ell)}{2}}   \otimes_\ell | \alpha_\ell \rangle \otimes_\ell \langle \beta_\ell| .
\ele(Qm)
By (\req(sum)), $Q(s)$ is a $s$-polynomial operator of degree $L$ with $Q(0)= {\rm Id}$ and the leading coefficient = the 
spin-inversion operator $R$. Hence $[R, Q(s)]=0$. Furthermore one can show $Q(-s) = S Q(s) S$ where $S = \bigotimes_{\ell} \sigma_\ell^{z}$. Indeed with $\tau_{i, j}$ in (\req(Sij)), we define $\widetilde{\tau}_{i, j} = \pm \tau_{i, j}$ according to $j-i \equiv \pm 1 \pmod{N}$. Replacing $\tau_{i, j}$ by $\widetilde{\tau}_{i, j}$ in the definition of ${\sf Q}_R$ in subsection \ref{ssec. Q}, we obtain the operator $\widetilde{\sf Q}_R$, hence $\widetilde{Q}_R (s) := s^{\frac{L}{2}} \widetilde{\sf Q}_R (s)$. The operators $Q_R$ and $\widetilde{Q}_R$ are related by $Q_R(-s) = S \widetilde{Q}_R (s) $, which implies 
$$
Q(-s) = Q_R(-s)Q_R(0)^{-1} = S \widetilde{Q}_R (s) \widetilde{Q}_R (0)^{-1} \widetilde{Q}_R (0)Q_R(0)^{-1} = S Q(s) \widetilde{Q}_R (0)Q_R(0)^{-1} .
$$
Setting $s=0$ in the above equality, one obtains  $\widetilde{Q}_R (0)Q_R(0)^{-1}= S$, therefore 
$Q(-s) = S Q(s) S$. 
Note that as an operator of $V$, one can write the $Q(s)$ in (\req(Qm)) in the form 
$$ 
Q (s)  =  s^{\frac{L}{2}} \sum_{  \alpha_\ell, \beta_\ell = \pm 1 } 
q^{\frac{  \sum_{ 1 \leq s < \ell \leq  L }  (\beta_\ell \alpha_s - \alpha_\ell \beta_s) }{4} }  s^{\frac{- \sum_\ell \alpha_\ell \beta_\ell }{2}}   \otimes_\ell | \alpha_\ell \rangle \otimes_\ell \langle \beta_\ell| .
$$
Up to $s^{\frac{L}{2}}$-factor, the above operator coincides with the six-vertex $Q$-operator (101) in \cite{B73} which holds for a generic $q$ with $\sum_{\ell} \alpha_\ell = \sum_{\ell} \beta_\ell =0$. Hence, (\req(Qm)) can be considered as the Baxter's classical six-vertex $Q$-operator in the root-of-unity case.

We are going to show the above $Q$-operator satisfies one, hence all by Theorem \ref{thm:QQ-Q}, of relations in (\ref{QQ0}) and (\ref{6Q0}) for cases of $L \leq 6$ by computational methods. For simple notation, we shall also write basis elements of the vector space $V$ in (\req(V)) by
\be
|\alpha_1, \cdots, \alpha_L  \rangle := \otimes_\ell | \alpha_\ell \rangle , \ \ \  \langle \beta_1, \cdots, \beta_L| := \otimes_\ell \langle \beta_\ell| .
\ele(base)

First, we consider the case  $L=2$, where $V$ is the 2-dimensional space with the basis, $|\alpha, -\alpha  \rangle$ ($\alpha = \pm 1$). Then $T(z)$ in (\req(6Tz)), $T^{(2)}$ and $Q$ are expressed by
$$
\begin{array}{ll}
T(z) =&(z+z^{-1}) (q^{S^z}+ q^{-S^z}) - 2(q^{\frac{-\sigma^z }{2}} \otimes q^{\frac{\sigma^z }{2}} + q^{\frac{\sigma^z }{2}} \otimes 
q^{\frac{-\sigma^z }{2}}) +(q-q^{-1})^2 (\sigma_- \otimes \sigma_+ + \sigma_+ \otimes \sigma_-),  \\
T^{(2)}( \widetilde{t}) = 
&  2(z^2+ 1 - z (q+q^{-1})) ( |1 , -1\rangle  \langle 1, -1| + |-1,1 \rangle  \langle -1 , 1 |\\
&+ z (q-q^{-1})^2 (|-1 ,1 \rangle  \langle 1 , -1 | + |1 ,|-1 \rangle \langle -1,1 |),  \\
Q (s) = &   \sum_{\alpha, \beta = \pm 1 }  z^{\frac{1- \alpha \beta}{2} } |\alpha , -\alpha \rangle \langle \beta, - \beta| . 
\end{array}
$$
Note that $Q(s)$ is a $z$-polynomial operator. The eigenvectors of $Q$ and $T^{(2)}$ are: $|1,-1  \rangle   + |-1,1  \rangle$ , $|1,-1  \rangle - |-1,1  \rangle$, with the $Q$- eigenvalue $(1+z)$ ,$(1-z)$, and $T^{(2)}$-eigenvalue $ 2(z^2+ 1 -z (q+q^{-1}))+z(q-q^{-1})^2$, $2(z^2+ 1 -z(q+q^{-1}))-z(q-q^{-1})^2$, respectively. The above $Q$-eigenvalues 
are indeed the polynomial $F(\tilde{t})$ for the Bethe equation (\req(6FBe0)), in which case the $Q$-functional equation (\ref{6Q0}), equivalent to (\req(P)), automatically holds.

For $L=4$, $V$ is of dimension 6 with the basis, $v_{j, k}, 1 \leq j < k \leq 4$, where $v_{j, k} = \otimes | \alpha_\ell \rangle$ with $\alpha_\ell = 1$ for $\ell=j, k$, and $-1$ otherwise. Denote $v_{j, k}^+ = v_{j, k} + R v_{j, k}$, $ v_{j, k}^- = v_{j, k} - R v_{j, k}$.  The matrix form of $Q$ for the basis $\{ v_{1, 2}, v_{3, 4}, v_{1, 3}, v_{2, 4}, v_{1, 4},  v_{2, 3} \}$ is 
$$
Q (s) =  \left( \begin{array}{cccccc}
         1& z^2 &z&z&qz& q^{-1}z  \\
         z^2& 1& z &z&q^{-1}z&qz  \\
         z& z& 1& z^2 &z&z  \\
         z&z& z^2& 1& z &z \\
         q^{-1}z&qz&z& z& 1& z^2  \\
         qz&q^{-1}z&z&z & z^2& 1  \\
\end{array} \right) , \ \ z=s^2.  
$$
Hence with respect to the basis $\{ v_{[1, 2]}^+,  v_{[1, 3]}^+,  v_{[1, 4]}^+, v_{[1, 2]}^-,  v_{[1, 3]}^-,  v_{[1, 4]}^- \}$, $Q$ is expressed by 
\be
Q(s) =   \left( \begin{array}{cccccc}
         1+z^2 &2z& (q+q^{-1})z&0&0&0\\
         2z & 1+z^2& 2z&0&0&0\\
         (q+q^{-1})z&2z &1+z^2&0&0&0 \\
         0& 0&0&1-z^2& 0&(q-q^{-1})z\\
         0& 0&0&0&1-z^2&0\\
         0& 0&0&(q^{-1}-q)z&0&1-z^2
\end{array} \right) ,   
\ele(QN3L4)
and the transfer matrix $T(z)$ is given by 
$$
T(z) = \left( \begin{array}{cccccc}
         2a^2b^2     &2abc^2& (a^2c^2 + b^2c^2)&0&0&0\\
         2abc^2      & (2a^2b^2 +c^4)& 2abc^2&0&0&0\\
         (a^2c^2+b^2c^2)&2abc^2 &2a^2b^2 &0&0&0 \\
         0& 0&0&2a^2b^2& 0&(a^2c^2- b^2c^2)\\
         0& 0&0&0&(2a^2b^2-c^4)&0\\
         0& 0&0&( b^2c^2-a^2c^2 )&0&2a^2b^2
\end{array} \right) ,   
$$
where $a, b, c$ are in (\req(6abc)). The eigenvectors for $T^{(2)}$ and $Q$ are: $v_{1,2}^+ - v_{1,4}^+, v_{1,2}^+ + v_{1,4}^+ - \frac{1}{4}(q+q^{-1} \pm \sqrt{q^2+q^{-2}+34})v_{1,3}^+$, $v_{1,2}^- \pm {\rm i}v_{1,4}^- , v_{1,3}^-$, with distinct eigenvalues. 
Indeed the $Q$-eigenvalues, 
$1+z^2 - (q+q^{-1})z$ , $1+z^2 + \frac{q+q^{-1} \mp \sqrt{q^2+q^{-2}+34}}{2}z$, $1-z^2 \pm {\rm i} (q-q^{-1})z$ , $1-z^2$,
are $F$-polynomials for the Bethe equation  (\req(6FBe0)) with $m=2$, (true as well for an arbitrary $q$ which is not necessary a root of unity). Then the $z$-polynomial expression of $Q$ yields the $Q$-functional equation (\ref{6Q0}). In fact, one can  directly verify (\ref{6Q0}) by using the matrix form of $Q(z)$. For example for $N=3$, the $Q$-functional equation (\ref{6Q0}) of size $L$ is equivalent to the following one:  
\be
3 Q(-s)Q(qs)Q(q^2 s) =  (1-z)^L Q(s)+ (1-qz)^L Q(q^2 s)  + (1-q^2 z)^L Q(q s)  . 
\ele(QfN3)
Then the matrix (\req(QN3L4)) satisfies the above relation for $L=4$.

In the cases $L=2, 4$, all $T^{(2)}$-eigenvalues are non-degenerated, and the $Q$-eigenvalues are the $z$-polynomial Bethe solutions of (\req(6FBe0)). This kind of relation between  $Q$-eigenvalues and Bethe solutions indeed holds for all non-degenerated $T^{(2)}$-eigenvalues for a general $L$  by assuming Bethe ansatz of the six-vertex model.  
\begin{thm}\label{thm:ndQBe} 
Let $T^{(2)}( \widetilde{t})$ be a non-degenerated $T^{(2)}$-eigenvalue for an arbitrary (even) $L$. Then its eigenspace is spanned by a vector in $V$ with $S^z=0$, and the corresponding $Q$-eigenvalue is a $z$-polynomial which defines a  Bethe solution $F(\widetilde{t})$ of  $(\req(6FBe0))$ with $m= \frac{L}{2}$. As a consequence, the $Q$-eigenvalue satisfies the $Q$-functional relation $(\ref{6Q0})$. 
\end{thm}
{\it Proof.} By Bethe ansatz, there exists a Bethe polynomial $F(\widetilde{t})$ for a solution of  $(\req(6FBe0))$ with $m \leq \frac{L}{2}$ so that $T^{(2)}(\widetilde{t})$ and $F(\widetilde{t})$ satisfy the relation (\req(TF0)). By Lemma \ref{lem:ndT}, $m= \frac{L}{2}$. This implies the one-dimensional eigenspace with the eigenvalue $T^{(2)}(z)$ has the basis element $v$ with $|S^z|= \frac{L}{2}-m =0$.  Then $Q(s)v = \lambda v$. Express $v$ as a linear combination of the standard basis, $
v = \sum_{\beta_1+ \ldots + \beta_L=0}  a_{\beta_1, \cdots , \beta_L} | \beta_1, \cdots , \beta_L \rangle$. We may assume one of  $a_{\beta_1, \cdots , \beta_L}$'s taking the value $1$, say $a_{\gamma_1, \ldots, \gamma_L}=1$, hence by $R v = \pm v$, $a_{-\gamma_1, \ldots, -\gamma_L}= \pm 1$. The relation (\req(Qm)) yields 
$$ 
\lambda = \langle \gamma_1, \ldots, \gamma_L | Q(s)v = \sum_{   \beta_1+ \ldots + \beta_L = 0 } a_{\beta_1, \cdots , \beta_L} 
q^{\frac{ - \sum_\ell  \gamma_\ell \sum_{s=1}^{\ell -1} \beta_s }{2} } q^{\frac{-\sum_\ell \gamma_\ell \beta_\ell}{4}} z^{\frac{\sum_\ell (1-\gamma_\ell \beta_\ell)}{4}}.
$$
In the above sum, $\frac{\sum_\ell (1-\gamma_\ell \beta_\ell)}{4}$ are integers  between $0$ and $\frac{L}{2}$, taking the value $0$ ( $\frac{L}{2}$) only when $\beta_\ell = \gamma_\ell$ ( $-\gamma_\ell $ resp.) for all $\ell$. By (\req(sum)), $\sum_\ell \gamma_\ell  \sum_{s=1}^{\ell-1} \gamma_s \equiv  \frac{-L}{2} \pmod{N}$, hence $\lambda \ (= \lambda (z))$ is a $z$-polynomial of degree $\frac{L}{2}$ with $\lambda (0)=1$ and the leading coefficient $=\pm 1$. Then the equation (\req(TF0)) holds with $F(\widetilde{t}):=\lambda (z)$.   
{\it Claim}. The roots of the polynomial $\lambda (z)$ do not contain a complete $N$-cyclic string, consequently, follows the result. Otherwise, by factoring out those roots of $\lambda (z)$ which form complete $N$-cyclic strings, one obtains a polynomial $F^\prime (\widetilde{t})$ with degree less than $\frac{L}{2}$, and roots containing no complete $N$-cyclic string, such that  $T^{2)}( \widetilde{t})$ and $F^\prime (\widetilde{t})$  again satisfy (\req(TF0)). Then $F^\prime (\widetilde{t})$ is a Bethe polynomial for a solution of $(\req(6FBe0))$ with $m < \frac{L}{2}$, which by Lemma \ref{lem:ndT} contradicts the non-degeneracy of $T^{(2)}( \widetilde{t})$. 
$\Box$ \par \vspace{.2in} 
As a corollary of the above theorem, the $Q$-functional relation holds when  all $T^{(2)}$-eigenvalues are non-degenerated.
\begin{prop}\label{prop:L2N} 
The necessary and sufficient condition for the $T^{(2)}$-matrix having no degenerated eigenvalue is the relation $\frac{L}{2} < N$. In this situation, the $Q$-operator $(\req(Qm))$ satisfies the $Q$-functional relation $(\ref{6Q0})$. 
\end{prop}
{\it Proof.} The condition $\frac{L}{2} < N$ is equivalent to $V$ being a subspace with the spin $S^z=0$, hence by Lemma \ref{lem:ndT}, equivalent to all $T^{(2)}$-eigenvalues being non-degenerated. Then the results follow from Theorem \ref{thm:ndQBe}. 
$\Box$ \par \noindent \vspace{.1in}
{\bf Remark.} The conclusion in the above proposition about non-degenerated $T$-eigenvalues and the corresponding eigenvalues of $Q$ holds for all $Q$-matrices, not a special property of the $Q$-operator (\req(Qm)). The computational argument we present here has further enhanced the correct form about the operator (\req(Qm)).  
$\Box$ \par \vspace{.2in}

We now discuss the case $L=6$. For $N \geq 5$, the $Q$-functional relation holds by Proposition \ref{prop:L2N}. So we need only to consider  the case $N=3$, where  $V$ of dimension 22 with the basis consisting of  
$v:= \otimes |1 \rangle , v':=  \otimes |-1 \rangle$  $v_{i,j,k}, 1 \leq i < j < k \leq 6$ (:= the vector in (\req(base)) with only the $i,j, k$-th factors being $|1 \rangle$), and the dual base,  $v^* , v^{'*}$, $v_{i,j,k}^{*}$s.

By Theorem \ref{thm:ndQBe}, there is only one degenerated $T^{(2)}$-eigenvalue, given by the 'pseudo-vacuum' $v$ with the eigenvalue $(1- \omega^{-1}\widetilde{t})^6 + (1-\widetilde{t})^6$. Denote the corresponding eigenspace by $W$, which is invariant under the $Q$-operator. So it remains to show that functional equation (\req(QfN3)) holds for $Q$ when restricting on $W$. With $F( \widetilde{t})=1$ in (\req(P)), one has the evaluation polynomial 
\be
P(\widetilde{t}) = 1- 20 \widetilde{t}^3+ \widetilde{t}^6 , 
\ele(PL6N3)
with the Drinfeld polynomial, $1- 20 x + x^2$ (see also (8.25) in \cite{De05}).
Denote the $S_R$-invariant vectors in $V$,
$$
u_0 := v_{1,3,5} + v_{2,4,6} , \ u_1 := \sum_{k=0}^5 v_{1_k, 2_k, 3_k} , \ w := \sum_{k=0}^5 v_{1_k,2_k,4_k} , \ w' := \sum_{k=0}^5 v_{1_k,2_k,5_k} ,
$$
where $i_k = \theta^k (i)$ with $\theta$ the cyclic permutation of $\{ 1, \cdots, 6 \}$ sending $k$ to $k+1$ by the identification $7=1$.  Then $R(u_i)=u_i \ (i=0,1)$, $R(w)= w'$.  The $sl_2$-loop-algebra structure of the space $W$ is determined by the values of the generators (\req(ST)) on $v, v'$, expressed by
\be
S^{-(3)}(v) (= T^{+(3)}(v')) = u_0 + u_1 + q w + q^2 w', \ ~ \
T^{-(3)}(v) (= S^{+(3)} (v') ) = u_0 + u_1 + q^2 w + q w', 
\ele(ST3)
with the relations $(S^{-(3)})^2(v) = 20 v'$, $(T^{-(3)})^2(v) = 20 v'$, $T^{-(3)} S^{-(3)} (v)  = S^{-(3)}T^{-(3)} (v) = 2 v'$. So $W$ is the 4-dimensional vector space with  
$\{ v, T^{-(3)}(v), S^{-(3)}(v), v' \}$ as a basis. We are going to determine the matrix form of $Q$ with respect to this basis. By (\req(Qm)), for any basis element $b$ of $V$ with $b^*$ as its dual element,  
$$
\langle b^* |Q(s)| b \rangle = 1, \ \ \langle b^* |Q(s)| (Rb) \rangle = z^3,
$$ 
and the following relations hold:
$$
\begin{array}{lll}
\langle \alpha_1, \ldots, \alpha_6 |Q(s)| v \rangle  &=   
q^{\frac{ - \sum_\ell \ell \alpha_\ell }{2} } z^{\frac{6-\sum_\ell \alpha_\ell }{4}}&=
z^{\frac{-\sum_\ell  \alpha_\ell}{2}} \langle v' |Q(s)| \alpha_1, \ldots, \alpha_6 \rangle  , \\
\langle \alpha_1, \ldots, \alpha_6 |Q(s)| v' \rangle  & =    
q^{\frac{ \sum_\ell \ell \alpha_\ell }{2} } z^{\frac{6+\sum_\ell \alpha_\ell }{4}} & = 
z^{\frac{\sum_\ell  \alpha_\ell}{2}} \langle v |Q(s)| \alpha_1, \ldots, \alpha_6 \rangle  .
\end{array}
$$
Hence,
$$
\begin{array}{ll}
\langle v^* |Q(s)|v \rangle = \langle v'^{*} |Q(s)|v' \rangle = 1, & \langle v'^{*} |Q(s)| v \rangle = \langle v^* |Q(s)| v' \rangle = z^3 , \\
\langle v_{1,3,5}^* |Q(s)|v \rangle = \langle v_{1,3,5}^* |Q(s)|v' \rangle = z^{\frac{3}{2}}, &
\langle v_{1,2,3}^* |Q(s)|v \rangle = \langle v_{1,2,3}^* |Q(s)|v' \rangle = z^{\frac{3}{2}}, \\
\langle v_{1,2,4}^* |Q(s)|v \rangle = \langle v_{1,2,5}^* |Q(s)|v' \rangle = q^2 z^{\frac{3}{2}}, &
\langle v_{1,2,5}^* |Q(s)|v \rangle = \langle v_{1,2,4}^* |Q(s)|v' \rangle = q z^{\frac{3}{2}}. 
\end{array}
$$
Since $[Q, S_R]=0$, and $v, v'$ are $S_R$-invariant vectors, the above relations yield
\bea(ll)
Q(s) v = v + z^3 v' + z^{\frac{3}{2}} u_0 + z^{\frac{3}{2}} u_1 + q^2 z^{\frac{3}{2}} w + q z^{\frac{3}{2}} w' &= v + z^{\frac{3}{2}} T^{-(3)}(v) + z^3 v' , \\
Q(s) v = z^3 v +  v' + z^{\frac{3}{2}} u_0 + z^{\frac{3}{2}} u_1 + q z^{\frac{3}{2}} w + q^2 z^{\frac{3}{2}} w' &= z^3 v + z^{\frac{3}{2}} S^{-(3)}(v) +  v' . \\
\elea(Qvv')
By
$$
\begin{array}{ll}
\langle v_{1,2,3}^* |Q(s)|v_{1,3, 5} \rangle = \langle v_{1,2,4}^* |Q(s)|v_{1,3, 5} \rangle = \langle v_{2,3,6}^* |Q(s)|v_{1,3, 5} \rangle =  z  , \\
\langle v_{1,2,3}^* |Q(s)|v_{2,4,6} \rangle = \langle v_{1,2,5}^* |Q(s)|v_{1,3, 5} \rangle = \langle v_{2,3,5}^* |Q(s)|v_{1,3, 5} \rangle = z^2  , \\
\end{array}
$$
the $S_R$-symmetry property of $Q$  yields 
\be
Q (s) (u_0) =  2z^{\frac{3}{2}} v + 2z^{\frac{3}{2}} v' + (1 + z^3)u_0 + (z  + z^2)(u_1 + w + w') .   
\ele(Qu0)
By 
$$
\begin{array}{ll}
\langle v_{1,3,5}^* |Q(s)|v_{1,2,3} \rangle  = \langle v_{1,2,4}^* |Q(s)|v_{1,2,3} \rangle  = z , &\langle v_{2,4,6}^* |Q(s)|v_{1,2,3} \rangle  = \langle v_{1,4,5}^* |Q(s)|v_{1,2,3} \rangle  = z^2 , \\
\langle v_{2,3,4}^* |Q(s)|v_{1,2, 3} \rangle = \langle v_{1,3,6}^* |Q(s)|v_{1,2, 3} \rangle =  q z   , & \langle v_{1,2,6}^* |Q(s)|v_{1,2,3} \rangle =  \langle v_{2,3,5}^* |Q(s)|v_{1,2,3} \rangle = q^2 z ,  \\
\langle v_{3,4,5}^* |Q(s)|v_{1,2,3} \rangle =  \langle v_{2,5,6}^* |Q(s)|v_{1,2,3} \rangle =  q z^2 , &
\langle v_{1,5,6}^* |Q(s)|v_{1,2,3} \rangle =  \langle v_{3,4,6}^* |Q(s)|v_{1,2,3} \rangle = q^2 z^2  , 
\end{array}
$$
the $R, S_R$-symmetries of $Q$ yield 
\be
Q (s) (u_1) = 6z^{\frac{3}{2}}v + 6 z^{\frac{3}{2}} v' + 3(z + z^2) u_0 + (1 - z - z^2 + z^3) u_1  .
\ele(Qu1)
Similarly, using
$$
\begin{array}{l}
\langle v_{1,3,5}^* |Q(s)|v_{1,2,4} \rangle  = \langle v_{4,5,6}^* |Q(s)|v_{1,2,4} \rangle  =
\langle v_{3,4,6}^* |Q(s)|v_{1,2,4} \rangle  = 
\langle v_{2,5,6}^* |Q(s)|v_{1,2,4} \rangle   = 
\langle v_{2,3,6}^* |Q(s)|v_{1,2,4} \rangle  \\ = z^2  , \\
\langle v_{2,4,6}^* |Q(s)|v_{1,2,4} \rangle  = \langle v_{1,2,3}^* |Q(s)|v_{1,2,4} \rangle   = \langle v_{1,4,5}^* |Q(s)|v_{1,2,4} \rangle = \langle v_{1,2,5}^* |Q(s)|v_{1,2,4} \rangle =\langle v_{1,3,4}^* |Q(s)|v_{1,2,4} \rangle \\ =  z, \\
\langle v_{2,3,4}^* |Q(s)|v_{1,2,4} \rangle = \langle v_{2,4,5}^* |Q(s)|v_{1,2,4} \rangle = q^2 z   , \ ~ \ \langle v_{1,2,6}^* |Q(s)|v_{1,2,4} \rangle =  \langle v_{1,4,6}^* |Q(s)|v_{1,2,4} \rangle =  qz,  \\
\langle v_{3,4,5}^* |Q(s)|v_{1,2,4} \rangle =  \langle v_{2,3,5}^* |Q(s)|v_{1,2,4} \rangle =  q^2 z^2 , \ ~ \
\langle v_{1,5,6}^* |Q(s)|v_{1,2,4} \rangle =  \langle v_{1,3,6}^* |Q(s)|v_{1,2,4} \rangle = qz^2  , 
\end{array}
$$
one obtains
\bea(ll)
Q(s) (w) &=  6q z^{\frac{3}{2}} v + 6q^2 z^{\frac{3}{2}} v' + 3(z + z^2) u_0 + (1 + z + z^2 ) w + (z+ z^2 +z^3 ) w' , \\
Q(s) (w') &=  6q^2 z^{\frac{3}{2}}v + 6q z^{\frac{3}{2}} v'  + 3(z  + z^2) u_0  +(z+ z^2 +z^3 ) w + (1 + z  + z^2 ) w' .
\elea(Qww')
By (\req(ST3)), (\req(Qu0)), (\req(Qu1)), and (\req(Qww')), we have 
\bea(ll)
Q(s)(T^{-3}(v)) &=  20 z^{\frac{3}{2}} v + 2 z^{\frac{3}{2}} v' + (1 + z^3) (u_0 +  u_1) +(q^2  + q z^3 ) w + (q  + q^2 z^3  ) w' , \\
&=20 z^{\frac{3}{2}}v + 2 z^{\frac{3}{2}} v' +  T^{-3}(v) + z^3S^{-3}(v) , \\
Q(s)(S^{-3}(v)) & = 2z^{\frac{3}{2}} v + 20 z^{\frac{3}{2}} v' + (1 + z^3) (u_0 +  u_1) +(q+ q^2z^3 ) w + (q^2 + qz^3) w' \\
& = 2z^{\frac{3}{2}} v + 20 z^{\frac{3}{2}} v' + z^3T^{-3}(v) +  S^{-3}(v).
\elea(QST)
Hence by (\req(Qvv')) and (\req(QST)),  with respect to the basis $\{ v, T^{-(3)}(v), S^{-(3)}(v), v' \}$ of $W$ the matrix form of $Q$ is
$$
Q (s)= \left( \begin{array}{cccc}
     1 &20 z^{\frac{3}{2}}  & 2z^{\frac{3}{2}}& z^3 \\
      z^{\frac{3}{2}} & 1 & z^3&0 \\
      0 & z^3& 1& z^{\frac{3}{2}}\\
     z^3& 2z^{\frac{3}{2}}&20 z^{\frac{3}{2}}& 1
\end{array} \right) , \ ~ \ s = z^{\frac{1}{2}} ,
$$
with the following eigenvectors and eigenvalues:
$$
\begin{array}{ll}
{\rm Eigenvector}, & {\rm Eigenvalue}  \\
3\sqrt{2}(v-v')+T^{-(3)}(v)-S^{-(3)}(v), & 1+3\sqrt{2}z^{\frac{3}{2}} -z^3 ; \\
3\sqrt{2}(v-v')-T^{-(3)}(v)+S^{-(3)}(v), & 1-3\sqrt{2}z^{\frac{3}{2}} -z^3  ; \\
\sqrt{22}(v+v')+T^{-(3)}(v)+S^{-(3)}(v),  &1+ \sqrt{22}z^{\frac{3}{2}} + z^3 ; \\
\sqrt(22)(v+v')-T^{-(3)}(v)-S^{-(3)}(v), & 1-\sqrt{22}z^{\frac{3}{2}} + z^3 . 
\end{array}
$$
Then the above $Q$-matrix satisfies the $Q$-functional equation (\req(QfN3)) for $L=6$. 
The evaluation polynomial (\req(PL6N3)) is related to the $Q$-eigenvalues by the factorizations:
$$
P(\tilde{t})=(1+3\sqrt{2}z^{\frac{3}{2}} -z^3)(1-3\sqrt{2}z^{\frac{3}{2}} -z^3) = (1+ \sqrt{22}z^{\frac{3}{2}} + z^3)(1-\sqrt{22}z^{\frac{3}{2}} + z^3) .
$$
This means  each $Q$-eigenvalue has its roots consisting of one square-root for both $10+3\sqrt{11}$ and $10-3\sqrt{11}$, which are zeros of the Drinfeld polynomial, as conjectured in section 3.2 of \cite{R05b}  about the relation  between roots of the $Q$-eigenvalue and the Drinfeld polynomial in the general situation. Note that the above example also shows that there are eigenvectors of $Q(s)$ which are not eigenvectors of the operator $S = \bigotimes_{\ell} \sigma_\ell^{z}$.  

\subsection{Mathematical verification of functional relations for a general $L$ in the root-of-unity six-vertex model \label{ssec. funL}}
In this subsection, we give a mathematical proof of the following theorem about functional relations of the six-vertex model at roots of unity.
\begin{thm}\label{thm:Qfun} 
The functional relations $(\ref{QQ0})$ and $(\ref{6Q0})$ hold for the $Q$-operator defined in $(\req(Qm))$.
\end{thm}
By Theorem \ref{thm:QQ-Q}, it suffices to show the relation (\ref{QQ0}) for $j=N$, which is equivalent to 
\be
N Q(s) Q(-s )   =  T^{(N)}(\omega \widetilde{t}).
\ele(QQN)
By (\req(Qm)), the matrix $NQ (s)Q (-s)$ on $V$ is given by
$$
\begin{array}{ll}
&s^{-L} \langle \alpha_1, \ldots, \alpha_L |N Q (s)Q (-s)| \beta_1, \ldots,  \beta_L \rangle   \\
=&  \sum_{k=0}^{N-1} \sum_{  \gamma_\ell = \pm 1  } q^{-k-\frac{1}{2} \sum_{\ell} \gamma_\ell}q^{\frac{ \sum_\ell  \gamma_\ell \sum_{i=1}^{\ell -1} (\alpha_i -\beta_i ) }{2} } q^{\frac{\sum_\ell \gamma_\ell(\alpha_\ell - \beta_\ell)}{4}} (-s)^{\frac{-\sum_\ell \gamma_\ell(\beta_\ell +\alpha_\ell )}{2}} (-1)^{\frac{\sum_\ell (1-\gamma_\ell \alpha_\ell)}{2}}\\
=& \sum_{k=0}^{N-1} \prod_{\ell=1}^L (-1)^{\frac{ 1- \alpha_\ell}{2}} ( q^{-k-\frac{1}{2} + \sum_{i=1}^{\ell -1} \frac{\alpha_i -\beta_i}{2}+ \frac{\alpha_\ell - \beta_\ell}{4}}  (-s)^{\frac{-(\alpha_\ell + \beta_\ell )}{2}} - q^{k+\frac{1}{2} - \sum_{i=1}^{\ell -1} \frac{\alpha_i -\beta_i}{2} - \frac{\alpha_\ell - \beta_\ell}{4}}  (-s)^{\frac{\alpha_\ell+ \beta_\ell }{2}}) 
\end{array}
$$
where $\sum_\ell \alpha_\ell \equiv \sum_\ell \beta_\ell \equiv 0  \pmod{N}$. Denote
\be
w_k( \alpha, \beta ; s) = (-1)^{\frac{ 1- \alpha}{2}} ( q^{-k-\frac{1}{2} + \frac{\alpha - \beta}{4}}  (-s)^{\frac{-(\alpha+ \beta )}{2}} - q^{k+\frac{1}{2} - \frac{\alpha - \beta}{4}}  (-s)^{\frac{\alpha+ \beta }{2}})  
\ele(wks)
for $ k \in \ZZ_N$, $\alpha, \beta = \pm 1$. Then we obtain the following entry-expression of the matrix  $NQ (s)Q (-s)$:
\bea(l) 
\langle \alpha_1, \ldots, \alpha_L |N Q (s)Q (-s)| \beta_1, \ldots,  \beta_L \rangle 
= s^L \sum_{k=0}^{N-1} \prod_{\ell=1}^L w_{k+  \sum_{i=1}^{\ell -1} \frac{\beta_i - \alpha_i}{2} }( \alpha_\ell , \beta_\ell; s) 
\elea(QQfor)

In the definition of six-vertex transfer matrix (\req(6Tz)), $L(s)$ in (\req(6VL)) is the $L$-operator with the $\CZ^2$-auxiliary space and the $\CZ^2$-quantum-space operator-entries, expressed in terms of $a(s), b(s), c$ in (\req(6abc)) with the relation $b(s)= a(qs)$. Similarly, 
the fusion matrix $T^{(j)} (\widetilde{t})$  can be obtained as the trace of $L^{(j)}(s)$, the fused $L$-operator  with the $\CZ^2$-quantum space and $\CZ^j$-auxiliary space associated to $L(s)$, where the $\CZ^j$-auxiliary space is the space of completely symmetric $(j-1)$-tensors of the $\CZ^2$-auxiliary space with the projection $P^{(j)}: \stackrel{j-1}{\otimes} \CZ^2 \rightarrow \CZ^j$. The $L^{(j)}$-operator, constructed from the Yang-Baxter solution $P^{(j)}(L(s) \otimes_{aux} L(qs)\otimes \cdots \otimes_{aux} L(q^{j-2}s))$ on the $\CZ^j$-auxiliary space, has been extensively studied in the six-vertex model \cite{Ka, Krp, KRS, Nep, Sk}. In order to exhibit the $T^{(j)}$-operator discussed in this paper, we now employ the fusion construction to compute the explicit form of $L^{(j)}(s)$. 
Using the canonical basis $\{ e^{(j)}_k \}^{j-1}_{k=0}$ of  $\CZ^j$-auxiliary space associated to the $\CZ^2$-basis $e_{\pm} = | \pm 1 \rangle $, and its dual basis $\{ e^{(j)*}_k \}^{j-1}_{k=0}$, 
$$
\begin{array}{l}
e^{(j)}_k := e_+^{j-1-k}e_-^{k} = \frac{1}{{j-1 \choose k}} \sum_{\beta_1+ \ldots+\beta_{j-1}=j-2k} | \beta_1, \ldots, \beta_{j-1} \rangle, \\
e^{(j)*}_k =  \sum_{\alpha_1+ \ldots+\alpha_{j-1}=j-2k} \langle \alpha_1, \ldots, \alpha_{j-1} |,
\end{array}
$$
one can write $L^{(j)}(s) = \bigg( L^{(j)}_{k, l} (s)\bigg)_{0 \leq k, l \leq j-1}$ where  $L^{(j)}_{k, l}(s)$ is the $\CZ^2$-(quantum-space) operator  expressed by 
$$
L^{(j)}_{k, l} (s) = \frac{\langle e^{(j)*}_k | L(s) \otimes_{aux} L(qs)\otimes_{aux} \cdots \otimes_{aux} L(q^{j-2}s) | e^{(j)}_l \rangle}{ \prod_{i =0}^{j-3} b( q^i s) } .
$$
Consider the auxiliary-space tensor $\CZ^j \otimes \CZ^2$ as a subspace of $\stackrel{j+1}{\otimes} \CZ^2$ with the identification
$$
e^{(j+1)}_{k+1} = \frac{1}{{j \choose k+1}} \bigg( {j-1 \choose k} e^{(j)}_{k} \otimes e_- +
{j-1 \choose k+1} e^{(j)}_{k+1} \otimes e_+ \bigg) , \ \ k=-1, \ldots, j-1. 
$$ 
Denote $
f^{(j-1)}_k := e^{(j)}_{k} \otimes e_- - e^{(j)}_{k+1} \otimes e_+$ for $ k=0, \ldots, j-2$.
Then $e^{(j+1)}_l, f^{(j-1)}_k$ form a basis of $\CZ^j \otimes \CZ^2$ with the dual basis $e^{(j+1)*}_l, f^{(j-1)*}_k$ given by
$$
\begin{array}{l}
e^{(j+1)*}_{k+1} = e^{(j)*}_{k} \otimes \langle -| \ + \ e^{(j)*}_{k+1} \otimes \langle +| , \\ f^{(j-1)*}_k = \frac{1}{{j \choose k+1}} \bigg( {j-1 \choose k+1} e^{(j)*}_{k} \otimes \langle -| \ - \ {j-1 \choose k} e^{(j)*}_{k+1} \otimes \langle +| \bigg) .
\end{array}
$$
One has
$$
L^{(j+1)}_{k, l}(s) = \langle e^{(j+1)*}_k | L^{(j+1)} (s) | e^{(j+1)}_l \rangle = \frac{1}{b( q^{j-2} s)} \langle e^{(j+1)*}_k | L^{(j)}(s) \otimes_{aux} L(q^{j-1}s) | e^{(j+1)}_l \rangle. 
$$
By induction argument, one can derive the explicit form of $L^{(j)}_{k, l}(s)$,  and its relation with the six-vertex $T^{(j)}$-operator in subsection \ref{ssec. TQfu}. Furthermore the fusion relation of $T^{(j)}$-operators follows from the equalities: 
$$
\begin{array}{l}
\langle e^{(j+1)*}_l | L^{(j)}(s) \otimes_{aux} L(q^{j-1}s) | f^{(j-1)}_k \rangle = 0 , \\
\langle f^{(j-1)*}_k | L^{(j)}(s) \otimes_{aux} L(q^{j-1}s) | f^{(j-1)}_l \rangle = b(q^{j-1}s) \langle e^{(j-1)*}_k | L^{(j-1)}(s)  | e^{(j-1)}_l \rangle .
\end{array}
$$
Therefore we obtain the following result:
\begin{prop}\label{prop:Lj} 
For an integer $j \geq 2$, the $\CZ^2$-operators $L^{(j)}_{k, l}(s) \ (0 \leq k, l \leq j-1)$  are zeros except $k- l = 0, \pm 1$, in which cases $L^{(j)}_{k, l} (= L^{(j)}_{k, l}(s))$ are expressed by
$$
L^{(j)}_{k, k} = \left( \begin{array}{cc}
     a(q^k s) &0\\
     0 & a( q^{j-1-k} s)   
\end{array} \right), \ \ L^{(j)}_{k+1, k} = \left( \begin{array}{cc}
     0 & q^{j-1-k} -q^{-j+1+k} \\
     0 & 0  
\end{array} \right), \ \ L^{(j)}_{k-1, k} = \left( \begin{array}{cc}
     0 & 0 \\
     q^k -q^{-k} & 0  
\end{array} \right).
$$
The fusion matrix $T^{(j)} (\widetilde{t})$ is given by  
\be
T^{(j)} (\widetilde{t}) = (sq^{\frac{j-2}{2}})^L {\rm tr}_{aux} ( \bigotimes_{\ell=1}^L L^{(j)}_\ell (s)), \ \ {\rm for} \ s \in \CZ , \ j \geq 2 ,
\ele(TjL)
which satisfy the fusion relations $(\req(Tfus))$ and $(\req(TfBdy))$ for $r=0$.
\end{prop} $\Box$ \par \noindent \vspace{.1in} 
{\bf Remark}. (1) Proposition \ref{prop:Lj}, except the boundary fusion relation $(\req(TfBdy))$ valid only for the $N$th root-of-unity $q$, holds as well for an arbitrary $q$.   

(2) The $L$-operator for the fusion matrix $T^{(j)} (\widetilde{t})$ is given by $sq^{\frac{j-2}{2}} L^{(j)}(s)$, whose entries are $s$-polynomials with the zero constant term $(sq^{\frac{j-2}{2}} L^{(j)}_{k,l}(s))_{|s=0}$ except the diagonal ones:
$$
(sq^{\frac{j-2}{2}} L^{(j)}_{k,k}(s))_{|s=0} = \left( \begin{array}{cc}
     - q^{-k+\frac{j-1}{2}} &0\\
     0 &  -q^{k-\frac{j-1}{2}}
\end{array} \right), \ 0 \leq k \leq j-1 .
$$
This implies the $V$-operator $T^{(j)} (0)= j \cdot {\rm Id_V}$. 
$\Box$ \par \vspace{.1in} 

We now show the $N$th $QQ$-relation (\req(QQN)) by using the explicit form of $L^{(N)}(s)$ in Proposition \ref{prop:Lj}. By (\req(TjL)) and $q^N=1$, $T^{(N)} (\omega \widetilde{t})$ is equal to the trace of the $L$th monodromy matrix for the local operator $s L^{(N)}(qs)$,
\be
T^{(N)} (\omega \widetilde{t})= s^L {\rm tr}_{\CZ^N} ( \bigotimes_{\ell =1}^L L^{(N)}_{\ell}(qs)).
\ele(TNom) 
By Proposition \ref{prop:Lj}, $L^{(N)}_{k, l}(qs)$ are zero except $k-l =0, \pm 1$:
\bea(c)
L^{(N)}_{k, k} (qs) = \left( \begin{array}{cc}
     w_k(1,1;s)  &0\\
     0 & w_k(-1,-1;s)   
\end{array} \right) , \\
L^{(N)}_{k, k-1}(qs)  = \left( \begin{array}{cc}
     0 & w_k(1,-1; s) \\
     0 & 0  
\end{array} \right), \ \
 L^{(N)}_{k, k+1}(qs)  = \left( \begin{array}{cc}
     0 & 0 \\
     w_k(-1,1; s) & 0  
\end{array} \right) ,
\elea(LNw)
where $w_k(\alpha, \beta;s)$ are functions defined in (\req(wks)). Note that the $w_k(\alpha, \beta;s)$ in the above expressions automatically determines the index $(k, l)$ of  $L^{(N)}_{k, l} (qs)$ where it appears, by the relation $(k, l)= (k, k+ \frac{\beta-\alpha}{2})$. For a given pair $\alpha, \beta = \pm 1$, the non-vanishing $L^{(N)}_{k, l} (qs)$ having an entry $w_k(\alpha, \beta;s)$ are indexed by $k \in \ZZ_N$.  
When evaluating the entry $\langle \alpha_1, \ldots, \alpha_L |  T^{(N)} (\omega \widetilde{t})| \beta_1, \ldots, \beta_L \rangle $ in (\req(TNom)), the non-zero contribution of $(L^{(N)}_{\ell})_{k_\ell, l_\ell}(qs)$ at the site $\ell$ is related to that at the first site $k:=k_1 \in \ZZ_N$ by the relation $k_\ell = k + \sum_{i=1}^{\ell-1} \frac{\beta_i-\alpha_i}{2}$. Hence using the $w_k(\alpha, \beta; s)$-expression of the non-zero $L^{(N)}_{k, l}(qs)$ in (\req(LNw)), one finds $\langle \alpha_1, \ldots, \alpha_L |  T^{(N)} (\omega \widetilde{t})| \beta_1, \ldots, \beta_L \rangle $ with the same expression as $\langle \alpha_1, \ldots, \alpha_L |  NQ(s)Q(-s)| \beta_1, \ldots, \beta_L \rangle$ in (\req(QQfor)). Therefore, we obtain the relation (\req(QQN)), then follows Theorem \ref{thm:Qfun}.

\section{Concluding Remarks}\label{sec. F}
As a parallel theory to the Onsager-algebra symmetry of the superintegrable CPM, we have derived the set of functional equations of the root-of-unity six-vertex model with the $Q$-operator encoding the loop-algebra symmetry of the model. By the similar construction of the $Q$-operator for the root-of-unity eight-vertex in \cite{B72}, we obtain the explicit form of $Q$-operator of the six-vertex model at roots of unity, and present the work in a way from the functional-relation aspect. As a check on our reasoning, we have first verified the $Q$-functional relation for the size $L$ from $2$ to $6$ by direct computations, then proceed to justify mathematically that the functional relations hold for the six-vertex $Q$-operator constructed here for a general $L$, much as in the case of superintegrable CPM. These results about the $Q$-operator of the six-vertex model are mainly derived from the mathematical structures it shares commonly with the chiral Potts transfer matrix in CPM. However a possible physical interpretation of the root-of-vertex six-vertex $Q$-operator is lacking at present. As suggested by the role of $Q$-operator in CPM, the problem appears to be an interesting one, and the answer could be valuable as well for the symmetry problems of other solvable lattice models, e.g., the root-of-unity eight-vertex model in \cite{FM04}. Note that in the study of the root-of-unity six-vertex model, the chain size $L$, the root's order $N$, and the spin $S^z$ can be arbitrary in general (e.g., in \cite{FM00, FM001, FM41}).
But, just to keep things simple, we have in this paper restricted our attention only to even $L$, odd $N$ and $S^z \equiv 0 \pmod{N}$, and leave possible generalizations to future work.

\section*{Acknowledgements}
The author is pleased to thank IHES in Bures sur Yvette, France, and Max-Planck-Institut f\"{u}r Mathematik in Bonn, Germany (October-November, 2005), MSRI and Department of Mathematics in U.C. Berkeley, USA (March-April, 2006) for the hospitality, where parts of this work were carried out.
This work is supported in part by National Science Council of Taiwan under Grant No NSC 94-2115-M-001-013.


\begin{thebibliography}{99}
\bibitem{AMP} G. Albertini, B. M. McCoy, and 
J. H. H. Perk, Eigenvalue spectrum of the
superintegrable chiral Potts model, in  Adv. Stud.
Pure Math., 19, Kinokuniya Academic (1989) 1--55.
%
\bibitem{B71} R. J. Baxter, Eight-vertex model in lattice statistics, Phys. Rev. Lett. 26 (1971) 832--833; One-dimensional anisotropic Heisenberg chain, Phys. Rev. Lett. 26 (1971) 834. 
%
\bibitem{B72} R. J. Baxter, Partition function of the eight vertex model, Ann. Phys. 70 (1972) 193--228.
%
\bibitem{B73} R. J. Baxter, Eight-vertex model in lattice statistic and  one-dimensional anisotropic Heisenberg chain I: Some fundamental eigenvalues,  II. Equivalence to a generalized Ice-type lattice model,  III. Eigenvalues of the transfer matrix and Hamiltonian, Ann. Phys. 76 (1973) 1--71.
%
\bibitem{Bax} R. J. Baxter, Exactly solved models
in statistical mechanics, Academic Press (1982).
%
\bibitem{B90} R. J. Baxter, Chiral Potts model: eigenvalues of the transfer matrix, Phys. Lett. A 146 (1990) 110--114.
%
\bibitem{B93} R. J. Baxter, Chiral Potts model with skewed boundary conditions, J.
Stat. Phys. 73 (1993) 461--495.
%
\bibitem{B94} R. J. Baxter, Interfacial tension of the chiral Potts model, J. Phys. A: Math. Gen. 27 (1994) 1837--1849.
%
\bibitem{B04} R. J. Baxter, The six and eight-vertex models revisited, J. Stat. Phys. 114 (2004) 43--66; cond-mat/0403138.
%
\bibitem{B049} R. J. Baxter, Transfer matrix functional relation for the generalized $\tau_2(t_q)$ model, J. Stat. Phys. 117 (2004) 1--25;  cond-mat/0409493.
%
\bibitem{BBP} R. J. Baxter, V.V. Bazhanov and
J.H.H. Perk,  Functional relations for transfer
matrices of the chiral Potts model, Int. J. Mod.
Phys. B 4 (1990) 803--870.
%
\bibitem{BazS} V.V. Bazhanov and Yu.G. Stroganov, Chiral
Potts model as a descendant of the six-vertex model, J.
Stat. Phys. 59 (1990) 799--817.
%
\bibitem{Be} H. A. Bethe, Zur theorie der metalle I. Eigenverte und eigenfunctionen der linearen atomkette, Z. Physik {\bf 71} (1931) 206--226.  
%
\bibitem{DFM} T. Deguchi, K. Fabricius and B. M. McCoy, The $sl_2$ loop algebra symmetry for the six-vertex model at roots of unity, J. Stat. Phys. 102 (2001) 701--736; cond-mat/9912141. 
%
\bibitem{De05} T. Deguchi: XXZ Bethe states as highest weight vectors of the 
$sl_2$ loop algebra at roots of unity, cond-mat/0503564. 
%
\bibitem{FM00} K. Fabricius and B. M. McCoy, Bethe's equation is incomplete for the XXZ model at roots of unity, J. Stat. Phys. 103 (2001) 647--678; cond-mat/0009279.
%
\bibitem{FM001} K. Fabricius and B. M. McCoy, Completing Bethe equations at roots of unity, J. Stat. Phys. 104 (2001) 573--587; cond-mat/0012501.
%
\bibitem{FM01} K. Fabricius and B. M. McCoy, Evaluation parameters and Bethe roots for the six vertex model at roots of unity, eds. M. Kashiwara and T. Miwa, {\it Progress in Mathematical Physics} Vol 23, Birkhauser Boston (2002), 119--144; cond-mat/0108057.
%
\bibitem{FM04} K. Fabricius and B. M. McCoy, Functional equations and fusion matrices for the eight vertex model, Publ. RIMS, 40 (2004) 905--932; cond-mat/0311122.
%
\bibitem{FM41} K. Fabricius and B. M. McCoy, Root of unity symmetries in the 8 and 6 vertex models, cond-mat/0411419.
%
\bibitem{Ka} M. Karowski, On the bound state problem in 1+1 dimensional field theories, Nucl. Phys. B 153 (1979) 244--252.
%
\bibitem{Krp} I. G. Korepanov, Hidden symmetries in the 6-vertex model of statistical physics, Zap. Nauchn. Sem. S-Petersburg, Otdel. Mat. Inst. Stekelov (POMI) 215 (1994) 163--177; hep-th/9410066.
%
\bibitem{Kor} C. Korff, Representation theory and Baxter's TQ equation for the six-vertex model. A pedagogical overview, cond-mat/0411758. 
%
\bibitem{KRS} P. P. Kulish, N. Yu. Reshetikhin and E. K. Sklyanin, Yang Baxter equation and representation theory, Lett. Math. Phys. 5 (1981) 393--403.
%
\bibitem{MR} B. M. McCoy and S. S. Roan, Excitation spectrum and phase structure of the chiral Potts model. Phys. Lett. A 150 (1990) 347--354.
%
\bibitem{Nep} R. I. Nepomechie, Solving the open XXZ spin chain with nondiagonal boundary terms at roots of unity, Nucl. Phys. B622 (2002) 615--632; hep-th/0110116.
%
\bibitem{R04} S. S. Roan, Chiral Potts rapidity curve descended from six-vertex model and symmetry group of rapidities, J. Phys. A: Math. Gen. 38 (2005) 7483--7499; cond-mat/0410011.
%
\bibitem{R05o} S. S. Roan, The Onsager algebra symmetry of $\tau^{(j)}$-matrices in the superintegrable chiral Potts model, J. Stat. Mech. (2005) P09007; cond-mat/0505698.
%
\bibitem{R05b} S. S. Roan, Bethe ansatz and symmetry in superintegrable chiral Potts model and root-of-unity six-vertex model, cond-mat/0511543.
%
\bibitem{Sk} E. K. Sklyanin, Some algebraic structures connected with the Yang-Baxter equation, Funct. Anal. Appl. 16 (1983) 263--270. 
\end{thebibliography}
\end{document}